\documentclass[
reprint,
floatfix,
 nofootinbib,
 amsmath,amssymb,
 aps,
 prb,
]{revtex4-2}

\usepackage{graphicx}% Include figure files
\usepackage{dcolumn}% Align table columns on decimal point
\usepackage{bm}% bold math
\usepackage{textcomp} % for textdegree
\usepackage{mathtools} % for dcase
\usepackage{braket} % for quantum state
\usepackage{color}

\def\nn{\nonumber}

\def\Qalgo{\cite{
1994-Shor,
1996-Grover,
1996-Ekert-Jozsa,
1999-Shor}}
\def\QECcode{\cite{1997-Kitaev-Yu,
2003-Kitaev-Yu,
2006-Raussendorf-et-al,
2007-Raussendorf-et-al,
2007-Raussendorf-2-et-al,
2011-Wang-et-al,
2012-Fowler-et-al,
2022-Tomaru-et-al}}
\def\spinqubit{\cite{
1998-Loss-et-al,
1999-Burkard-et-al,
2013-Zwanenburg-et-al,
2023-Burkard-et-al}}
\def\onequbit{\cite{
2014-Veldhorst-et-al,
2015-Laucht-et-al,
2015-Muhonen-et-al,
2016-Kawakami-et-al,
2018-Yoneda-et-al}}
\def\twoqubit{\cite{
2022-Noiri-et-al,
2022-Xue-et-al,
2022-Mills-et-al}}
\def\manufacting{\cite{
2016-Maurand-et-al,
2017-Veldhorst-et-al,
2021-Xue-et-al,
2020-Li-et-al,
2021-Gonzalez-et-al,
2022-Ruffino-et-al,
2022-Zwerver-et-al,
2023-Elsayed-et-al,
2024-Neyens-et-al,
2024-Stuyck-et-al}}
\def\coherence{\cite{
2014-Veldhorst-et-al,
2020-Struck-et-al,
2022-Stano-et-al,
2023-Tanttu-et-al,
2024-Kuno-et-al}}
\def\shuttlingQEC{\cite{
2019-Boter-et-al,
2022-Boter-et-al,
2024-SpinBus,
2024-Siegel-et-al,
2025-Siegel-et-al}}
\def\conveyorall{\cite{
2020-Buonacorsi-et-al,
2022-QuBus,
2023-Langrock-et-al,
2024-QuBus,
2024-QuBus-EPR,
2024-Losert-et-al,
2024-QuTech,
2024-Jeon-Benjamin-Fisher,
2024-David-et-al,
2024-Oda-et-al,
2024-Nemeth-et-al}}
\def\conveyorexp{\cite{
2022-QuBus,
2024-QuBus,
2024-QuBus-EPR,
2024-QuTech}}
\def\conveyorexpp{\cite{
2022-QuBus,
2023-Langrock-et-al,
2024-QuBus,
2024-QuBus-EPR,
2024-QuTech}}
\def\bucketall{\cite{
2018-Zhao-Hu,
2019-Mills-et-al,
2020-Ginzel-et-al,
2021-Yoneda-et-al,
2022-Noiri-et-al,
2023-QuTech,
2024-Krzywda-et-al}}

\begin{document}

%\preprint{APS/123-QED}

\title{Digital-Controlled Method of Conveyor-Belt Spin Shuttling\\ in Silicon for Large-Scale Quantum Computation}

\author{Ryo Nagai}
\email{ryo.nagai.jd@hitachi.com}
\affiliation{
Center for Exploratory Research, 
Research and Development Group, 
Hitachi, Ltd., 
Kokubunji, Tokyo, 185-8601, Japan.
}

\author{Takashi Takemoto}
\affiliation{
Center for Exploratory Research, 
Research and Development Group, 
Hitachi, Ltd., 
Kokubunji, Tokyo, 185-8601, Japan.
}

\author{Yusuke Wachi}
\affiliation{
Center for Exploratory Research, 
Research and Development Group, 
Hitachi, Ltd., 
Kokubunji, Tokyo, 185-8601, Japan.
}

\author{Hiroyuki Mizuno}
\affiliation{
Center for Exploratory Research, 
Research and Development Group, 
Hitachi, Ltd., 
Kokubunji, Tokyo, 185-8601, Japan.
}

\date{\today}

\begin{abstract}
We propose a digital-controlled conveyor-belt shuttling method for silicon-based quantum processors, addressing the scalability challenges of conventional analog sinusoidal implementations. By placing a switch matrix and low-pass filters in a cryogenic environment, our approach synthesizes near-sinusoidal waveforms from a limited number of DC voltage levels. Simulation results demonstrate that the proposed method achieves fidelity comparable to analog methods while significantly reducing wiring overhead and power dissipation. Moreover, the design offers robustness against device-level variations, enabling large-scale integration of high-fidelity spin shuttling for quantum error correction.
\end{abstract}

\maketitle

%%%%%%%%%%%%%%%%%%%%%%%%%%%%%%%%%%%%%%%%%%%%%%%%%%%%%%%%%%%%%%%%%%%
\section{Introduction}
Quantum computers have attracted significant attention due to their potential for solving large-scale problems and for achieving computation speeds difficult to attain with classical machines~\Qalgo. However, current Noisy Intermediate-Scale Quantum (NISQ) devices suffer from high error rates, which makes it challenging to fully benefit from large-scale quantum algorithms. Consequently, quantum error correction (QEC) is considered critical for realizing the full potential of quantum computing, motivating the development of techniques to achieve high-fidelity and scalable quantum operations~\QECcode.

There are various physical implementations of quantum computers. 
In this paper, we focus on silicon-based quantum computers that
use the electron spin in silicon as a qubit~\spinqubit. This is because (i) they could take advantage of CMOS-compatible 
fabrication processes, potentially offering high scalability~\manufacting, (ii) their single-spin coherence times are known to be long~\coherence, 
and (iii) high-fidelity single-qubit~\onequbit~and two-qubit~\twoqubit~gates have 
already been experimentally demonstrated.

Realizing QEC in silicon spin qubits for large-scale 
quantum computation requires high-precision control of interactions among 
multiple qubits. In particular, as discussed in Refs.~\shuttlingQEC, a technique 
that preserves quantum coherence while shuttling electrons 
over distances of the order of $\mathcal{O}(10-100\,\mu\mathrm{m})$ is crucial in realistic 
architectures. Here, “shuttling” explicitly refers to physically transferring 
electrons within the device while maintaining their quantum state. Therefore, 
high-fidelity spin shuttling is one of the key technologies for QEC implementation in silicon quantum 
computers.

In silicon spin qubits, shuttling electrons physically can be largely classified into two approaches: \textit{bucket-brigade} shuttling \bucketall~and \textit{conveyor-belt} shuttling \conveyorall. Bucket-brigade shuttling transfers electrons successively between adjacent quantum dots and has the advantage of relatively simple control. However, for long-distance transport, the number of gates tends to increase, complicating the control system. In contrast, conveyor-belt shuttling continuously deforms a potential to transport electrons. This can be advantageous for long-distance motion because it can be implemented by repeatedly applying a certain control pattern over a distance. Indeed, the work of Langrock~\textit{et al.} \cite{2023-Langrock-et-al} suggests that conveyor-belt shuttling may be preferable for constructing large-scale architectures that realize QEC.

\begin{figure*}[t]
  \centering
  \includegraphics[width=17cm]{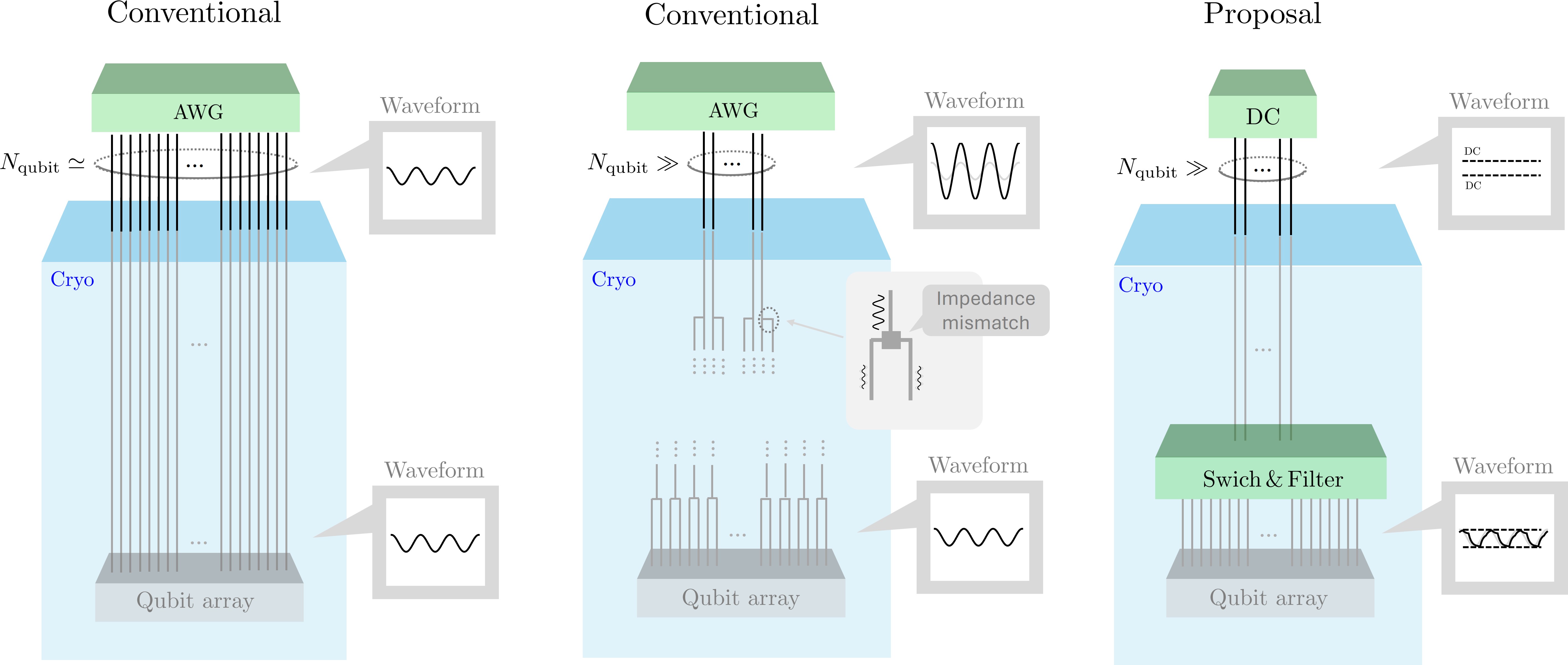}
  \caption{
  Conceptual figures of the analog-controlled approach (left and middle) and the proposed digital-controlled approach (right) for spin qubit shuttling in Si devices. 
  The conventional approach~\conveyorexp~corresponds to the middle figure,
  where a few (phase-modulated) sinusoidal waveforms are applied periodically to gate electrodes. This method is free from wiring problem at room temperature but 
  requires many fan-out cables and these would causes the amplitude dumping of waveforms due to the impedance mismatch. To mitigate the dumping effect, the larger voltage would be needed. This fan-out problem can be solved by introducing individual control depicted in the left figure, however, it will face wiring problem. 
  Our approach depicted the right figure has a potential to solve the dilemma. 
  We generate near (phase-modulated) sinusoidal waveform by placing the waveform-generation functionality (switches and filters) in the dilution refrigerator (Cryo).
  }
  \label{fig-original_vs_proposal}
\end{figure*}

However, from the perspective of large-scale integration, practical implementations of conveyor-belt shuttling would face challenges.  
A major reason is that current implementations primarily rely on analog control signals, such as phase-modulated or bias-tuned sinusoidal waveforms applied to gate electrodes~\conveyorexpp. Specifically, the proposed methods utilize a small set of AC voltages applied periodically to gate electrodes along the shuttling direction. 
The middle of Fig.~\ref{fig-original_vs_proposal} is a schematic figure of this method.
While such approaches work well for small-scale devices~\conveyorexp, scaling them up to large systems would require complex fanout cables to supply voltages to numerous shuttle gates, which can lead to waveform degradation due to impedance mismatch and signal distortion.
In particular, if the waveform amplitude is reduced due to such effects (roughly proportional to the inverse square root of the fan-out number), the quantum dot size would increase, potentially resulting in significant fidelity degradation. Therefore, to maintain high shuttling fidelity in the conventional method, one would need to compensate for the amplitude reduction by increasing the applied voltage beforehand, which would lead to unnecessary power consumption.

The fan-out problem described above can be mitigated by adopting a control scheme like that shown in the left of Fig. \ref{fig-original_vs_proposal}, which uses individual voltage control for each gate.
However, this approach requires running a large number of analog cables from room temperature to the cryogenic environment, with each quantum dot-sized control line, making large-scale implementation difficult.
From these considerations, we consider that the conventional analog control schemes face a fundamental trade-off between scalability and fidelity, which becomes a significant obstacle as system size increases.

To address this issue, we propose a new approach for implementing conveyor-belt shuttling in silicon qubits, replacing conventional analog control with a “digital-controlled” method (depicted by the right of Fig.~\ref{fig-original_vs_proposal}). We theoretically evaluate the shuttling fidelity of this approach and show that it can achieve fidelity comparable to the conventional analog method. Additionally, we highlight the advantages of this digital scheme in a large-scale quantum computing architecture, suggesting the feasibility of long-distance, high-fidelity spin shuttling compatible with QEC.

The paper is organized as follows{\footnote{
In this paper, we use natural units ($\hbar=c=1$) for simplicity, 
restoring physical units when necessary.
}}.
In Sec.~\ref{sec:proposal}, we introduce our proposed digital-controlled scheme 
and present simulations that highlight its advantages in fidelity and reduced wiring. 
Section~\ref{sec:conclusion} concludes the paper and provides an outlook on 
future research directions and applications of our method in fault-tolerant 
quantum computing architectures.
We provide review of the conventional analog approach and the detail of our numerical evaluations in appendices.

%%%%%%%%%%%%%%%%%%%%%%%%%%%%%%%%%%%%%%%%%%%%%%%%%%%%%%%%%%%%%%%%%%%

%%%%%%%%%%%%%%%%%%%%%%%%%%%%%%%%%%%%%%%%%%%%%%%%%%%%%%%%%%%%%%%%%%%
%%%%%%%%%%%%%%%%%%%%%%%%%%%%%%%%%%%%%%%%%%%%%%%%%%%%%%%%%%%%%%%%%%%
\section{Proposed Method}
\label{sec:proposal}
In this section, we propose a novel “digital-controlled” approach to conveyor-belt shuttling.
Section~\ref{sec:proposed-basic} explains the concept and a possible implementation. We then evaluate the method’s shuttling fidelity (Sec.~\ref{sec:proposed-fidelity}) and the power dissipation (Sec.~\ref{sec:proposed-power}) to assess its viability.

%------------------------------------------------------------------
\subsection{Basic structure}
\label{sec:proposed-basic}
In conventional shuttling methods, the arbitrary waveform generators (AWGs) are typically located outside the dilution refrigerator, which inevitably leads to scalability issues as described earlier.  
Our key insight is to move part of the waveform-generation circuitry (the switch matrix and simple analog filters) into the dilution refrigerator.  
A schematic figure of our proposal is shown on the right of Figure~\ref{fig-original_vs_proposal}.  
In this approach, only a small number of DC lines need to be supplied from room temperature, effectively eliminating wiring overhead.  
Furthermore, because the waveform amplitude is generated directly from DC voltages, our method allows for more reliable and stable amplitude control.  
This structure thus effectively overcomes the trade-off faced by conventional analog control schemes, enabling better scalability without sacrificing fidelity.
\begin{figure*}[t]
  \centering
  \includegraphics[width=13cm]{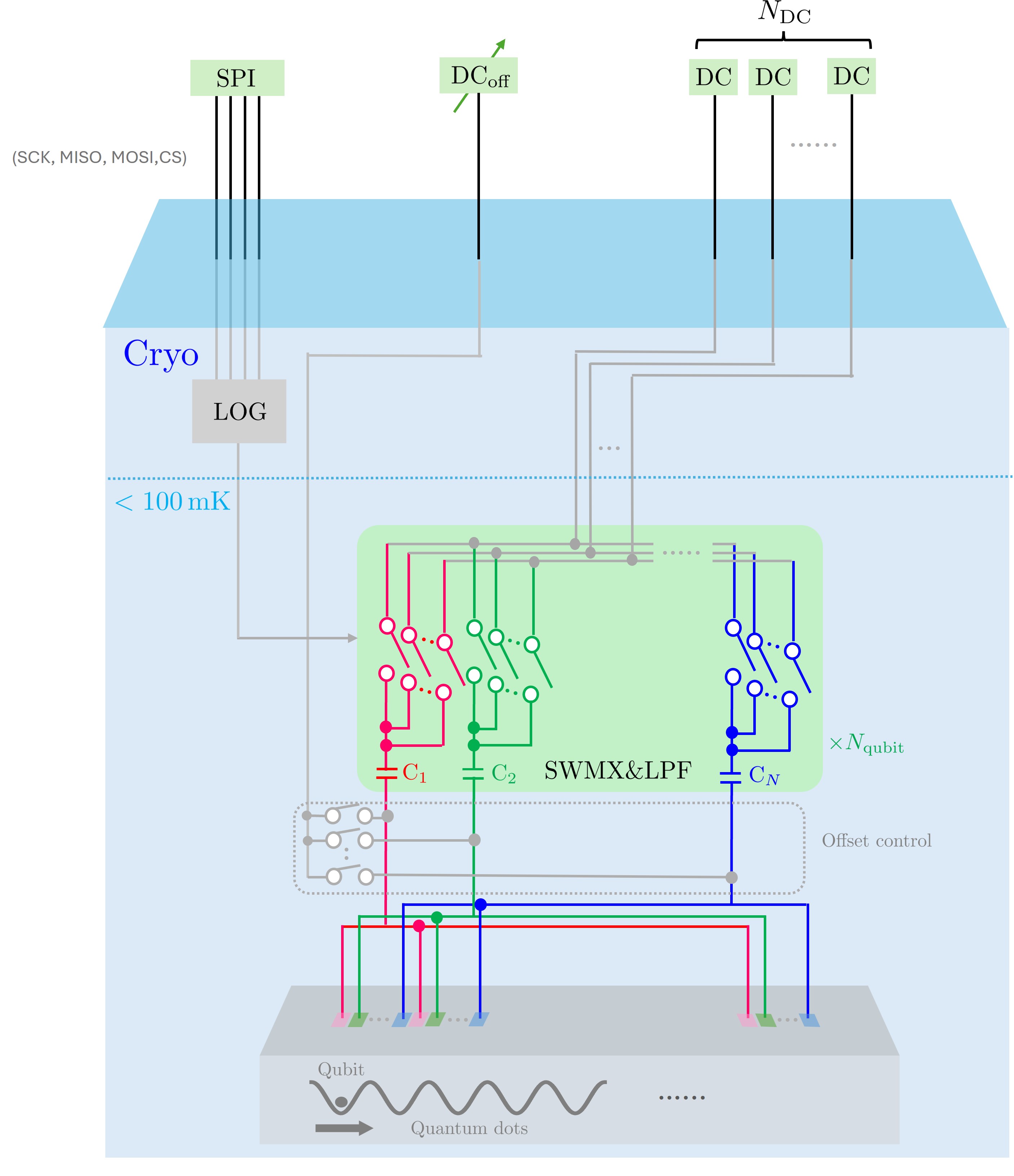}
  \caption{One example of the proposed implementation (see Fig.~\ref{fig-original_vs_proposal}, right). A switch matrix (SWMX), low-pass filter (LPF), and offset control are placed in the sub-100\,mK region of the dilution refrigerator (Cryo). Logic circuits (LOG) at a higher temperature stage control them. At room temperature, only $N_{\mathrm{DC}}$ DC voltage sources plus one additional serial control line (and a few other lines) are necessary, independent of $N_{\rm{qubit}}$.}
  \label{fig-structure}
\end{figure*}

\begin{figure*}[t]
  \centering
  \includegraphics[width=15cm]{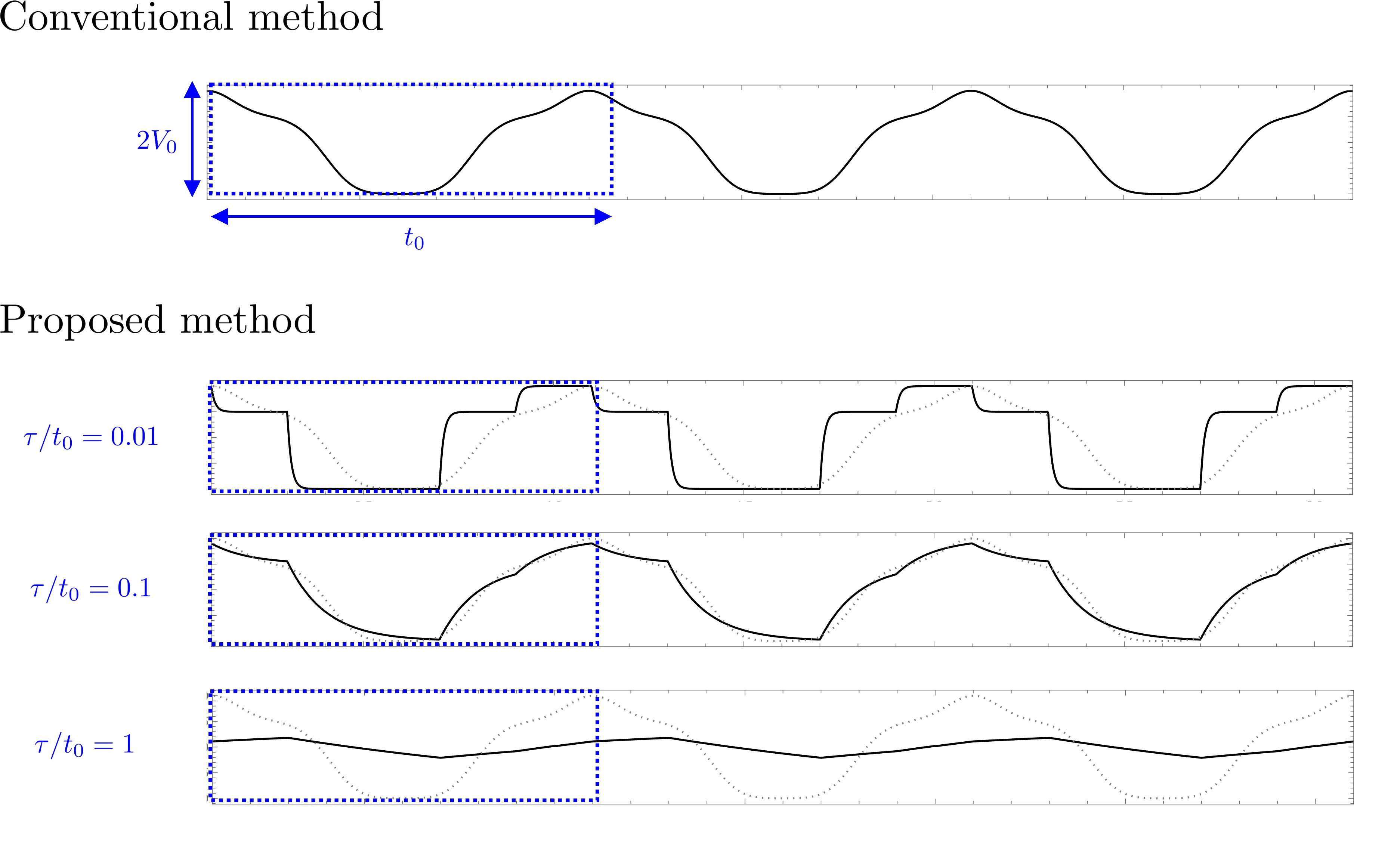}
  \caption{Effect of the time constant ($\tau$) on the voltage waveform in the proposed method. The top panel shows the original sinusoid; the bottom panels show resulting waveforms for $\tau/t_0=0.01,0.1,1.0$.}
  \label{fig-V-proposed}
\end{figure*}

Figure~\ref{fig-structure} depicts one possible concrete implementation of the above idea. At room temperature, we have $N_{\mathrm{DC}}(\ll N_{\rm{qubit}})$ DC voltage sources that generate a finite set of voltage levels. Inside the sub-$100\,m$K stage of the dilution refrigerator, a switch matrix (SWMX) and low-pass filter (LPF) are placed. The SWMX periodically switches between these $N_{\mathrm{DC}}$ levels with period $t_0$, thereby creating $N$ periodic rectangular waveforms{\footnote{We note that
increasing $N$ improves robustness but it also increases heat dissipation as we will discuss in section \ref{sec:proposed-power}.}}. In a system with $N_{\rm{qubit}}$ qubits, we might generate up to $N_{\rm{qubit}} N$ such waveforms, each of which is passed through RC filters that smear out the waveforms, creating near-sinusoidal shapes. The schematics of the smeared waveform is depicted in Fig.~\ref{fig-V-proposed} where $\tau$ denotes the time constant of LPF. 
The top of Fig.~\ref{fig-V-proposed} depicts the typical waveform in the conventional method.
We see that the our method can generate the similar waveform by adjusting the time constant of LPF.
These smeared waveforms are then distributed to $N_{\rm{qubit}} N_{\mathrm{gate}}$ gate electrodes, as shown schematically at the bottom of Fig.~\ref{fig-structure}. 
Here, $N_{\mathrm{gate}}$ denotes the number of gate electrodes required 
per shuttling process, 
which is typically around 40 in a QEC architecture~\cite{2022-Boter-et-al}.
Offset voltages can be individually adjusted by additional DC sources, correcting for device-to-device variations. Logic circuit (LOG) at a few-K stage controls switching and filtering parameters via a serial interface such as SPI.

This architecture directly addresses the previously stated challenges:
\begin{itemize}
\item 
As shown in Fig.~\ref{fig-structure}, the wiring requires only $N_{\rm{DC}}+1+4$ lines where $N_{\rm{DC}}$ can be much smaller than the number of qubit, $N_{\rm{qubit}}$. 
This overcomes the wiring challenge associated with scaling up quantum systems and ensures better scalability.

\item 
Generating high-frequency components near the quantum dots in the dilution refrigerator greatly reduces waveform distortion from cables and connectors, thereby minimizing the need for intricate calibration.
\end{itemize}

Our proposed method, as described earlier, effectively resolves the challenges faced by conventional methods. However, two concerns must be addressed to ensure its practical implementation.
The first is fidelity. Since the applied voltage waveforms differ from those in the conventional method, it is necessary to verify whether high fidelity can still be maintained. The second concern is power dissipation. Since the waveform generator is placed inside the dilution refrigerator, its power consumption must remain within the cooling capacity of the refrigerator. In the following sections, we analyze these two aspects through simulations and demonstrate that both concerns can be adequately addressed.

%------------------------------------------------------------------
\subsection{Shuttling Fidelity}
\label{sec:proposed-fidelity}

We first evaluate, via simulation, the shuttling fidelity of the proposed digital approach. For illustration, we take $N_{\mathrm{DC}}=3$, which minimizes the external wiring to three DC lines. (More DC lines would allow finer approximations of the analog waveforms and potentially higher fidelity.)

For $N_{\mathrm{DC}}=3$, the applied waveforms are given as follows. Three DC voltage sources generate $(V_0,\,V_0/2,\,-V_0)$. The SWMX cycles through them to produce the rectangular wave whose forms are given as
\begin{widetext}
\begin{align}
V_i(t)
&=
V\Bigl(t + \frac{N-i+1}{N}\,t_0\Bigr),
\\
V(t)
&=
\begin{cases}
 \tfrac{V_0}{2}, & 0 \le t \bmod t_0 \le \Delta t_1, \\
 -V_0,           & \Delta t_1 < t \bmod t_0 \le \Delta t_1+\Delta t_2, \\
 \tfrac{V_0}{2}, & \Delta t_1+\Delta t_2 < t \bmod t_0 \le \Delta t_1+\Delta t_2+\Delta t_3, \\
 V_0,            & \Delta t_1+\Delta t_2+\Delta t_3 < t \bmod t_0 \le t_0,
\end{cases}
\label{eq:Vpulse}
\end{align}
\end{widetext}
where $\Delta t_i$ is chosen to mimic a phase-modulated sinusoid. For the following analysis, we set $\Delta t_1 = \Delta t_3 = \Delta t_4 = t_0/5$ and $\Delta t_2=2t_0/5$. The resulting rectangular wave is then filtered by LPFs with a time constant $\tau$. 
As in the conventional method, the voltage applied to each gate is periodic, and the voltage applied to the-$(N+k)$th gate is the same as that applied to the $k$-th gate.

\begin{figure*}[t]
  \centering
  \includegraphics[width=16cm]{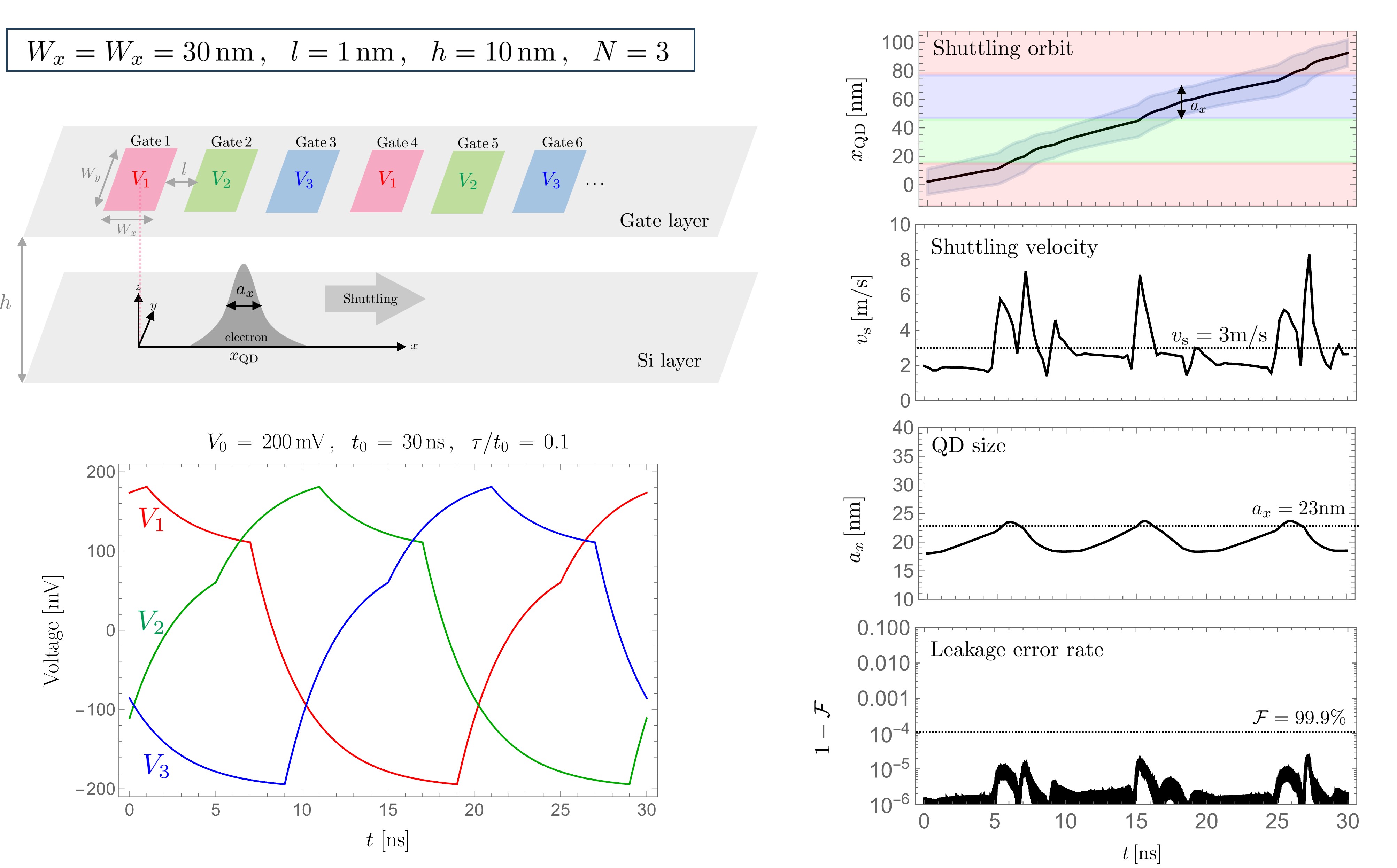}
  \caption{
  Numerical results for the proposed digital shuttling . 
  In the top left panels, we show the schematic picture of the device we considered. 
  Multiple gate electrodes are arranged along the shuttling direction (the $x$-axis) with a uniform spacing $l$. Each gate covers an area of ($W_x\times W_y$) and is positioned above the silicon layer at a distance $h$. 
  In the bottom left, the applied gate voltages is shown. 
  In our analysis, we set $W_x=W_y=30\,\mathrm{nm}, l=1\,\mathrm{nm}, h=10\,\mathrm{nm}, N=3, V_0=200\,\mathrm{mV}, 1/f=30\,\mathrm{ns}$, $E_{v,0}=200\,\mu\mbox{eV}$ and $\theta=0.3^\circ$. The right plots show the time evolution of the quantum dot position $x_{\rm{QD}}$, the shuttling velocity $v_{\rm{s}}$, the quantum dot size $a_x$, and the leakage error $1-\mathcal{F}$.}
  \label{fig-sim-proposed}
\end{figure*}
Let us estimate the shuttling fidelity. 
For a concrete example,
we consider the device shown in the left top of Fig.~\ref{fig-sim-proposed} with $W_x=W_y=W=30\,\mathrm{nm}$, $l=1\,\mathrm{nm}${\footnote{
The analysis for larger $l$ is provided in the Appendix \ref{app:delta-random}, where it is confirmed that the overall results remain largely unchanged.
}} and $h=10\,\mathrm{nm}${\footnote{
$h=10\,\rm{nm}$ is a typical value for SiMOS based silicon qubit device \cite{2024-Chen-et-al}. 
On the other hand, in Si/SiGe heterostructures, 
the typical separation $h$ is about 30\,$\mathrm{nm}$ \cite{2020-Lawrie-et-al,2025-Koch-et-al}.
To evaluate the impact of this difference, we also performed calculations assuming $h=30\,\mathrm{nm}$ in the Appendix.~\ref{app:delta-random}
}}.
The left bottom of Fig.~\ref{fig-sim-proposed} shows the applied gate waveforms (bottom), where $V_{(i\,\mathrm{mod}\,3)}$ denotes the gate voltage to the $i$th gate. The quantum dot size ($a_x$) and position ($x_{\rm{QD}}$) are computed from these waveforms. (See Appendix \ref{app:VQD} for the detail.) The right panels of Fig.~\ref{fig-sim-proposed} show the position of quantum dot $x_{\rm{QD}}$, shuttling velocity $v_{\rm{s}}$, the size of quantum dot $a_x$, and shuttling infidelity $1-\mathcal{F}$.
We estimate the shuttling infidelity assuming 
the valley splitting $E_{v,0}=200\,\mu\mathrm{eV}$ and the smooth interface with miscut angle of $\theta=0.3^\circ$. See Appendix \ref{app:delta} for the detail. 
We observe that the proposed method, employing smeared rectangular waveforms instead of sinusoidal ones, can achieve shuttling infidelity $1-\mathcal{F}$ below $10^{-4}$, which is comparable to the conventional approach (Fig.~\ref{fig-sim-conventional}).

Following the approach in references {\cite{
2022-Wuetz-Losert-Coppersmith-Friesen-et-al,
2023-Losert-Coppersmith-Friesen-et-al}}, 
we also performed calculations incorporating silicon interface roughness in Appendix \ref{app:delta-random}. Even under these more realistic conditions, the fidelity remains comparable to that of the conventional method.

\begin{figure*}[t]
  \centering
  \includegraphics[width=18cm]{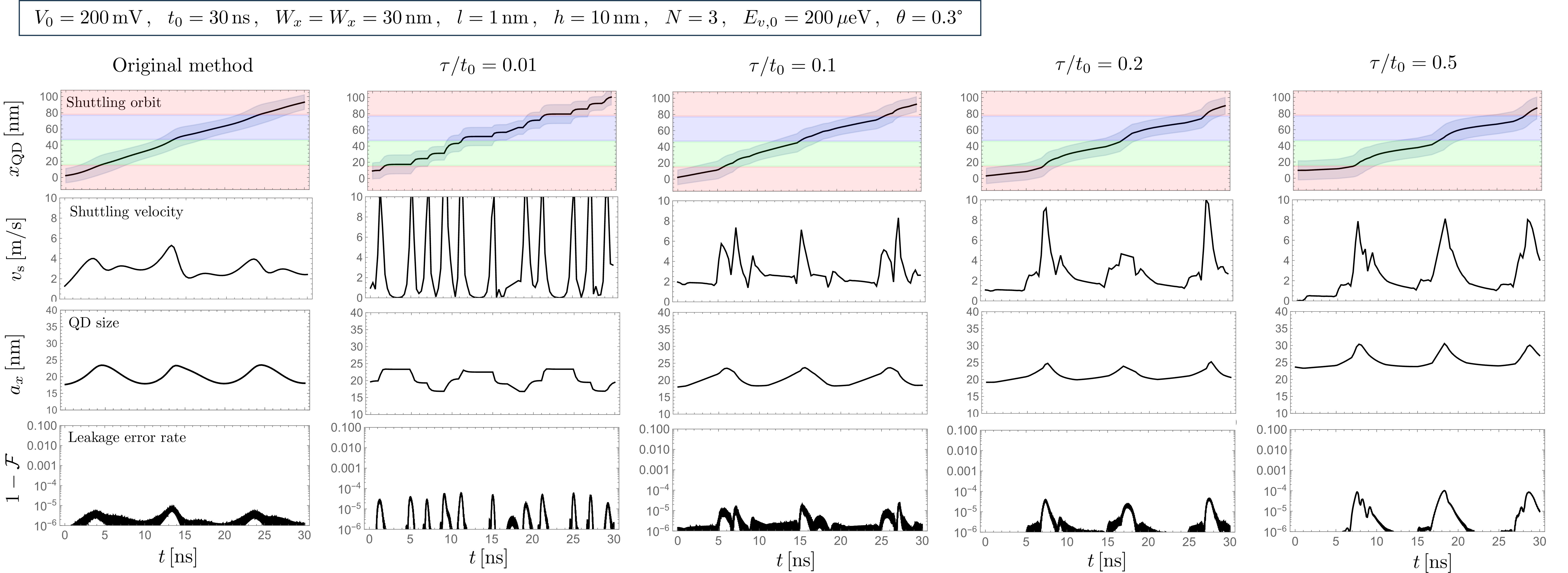}
  \\
  \includegraphics[width=18cm]{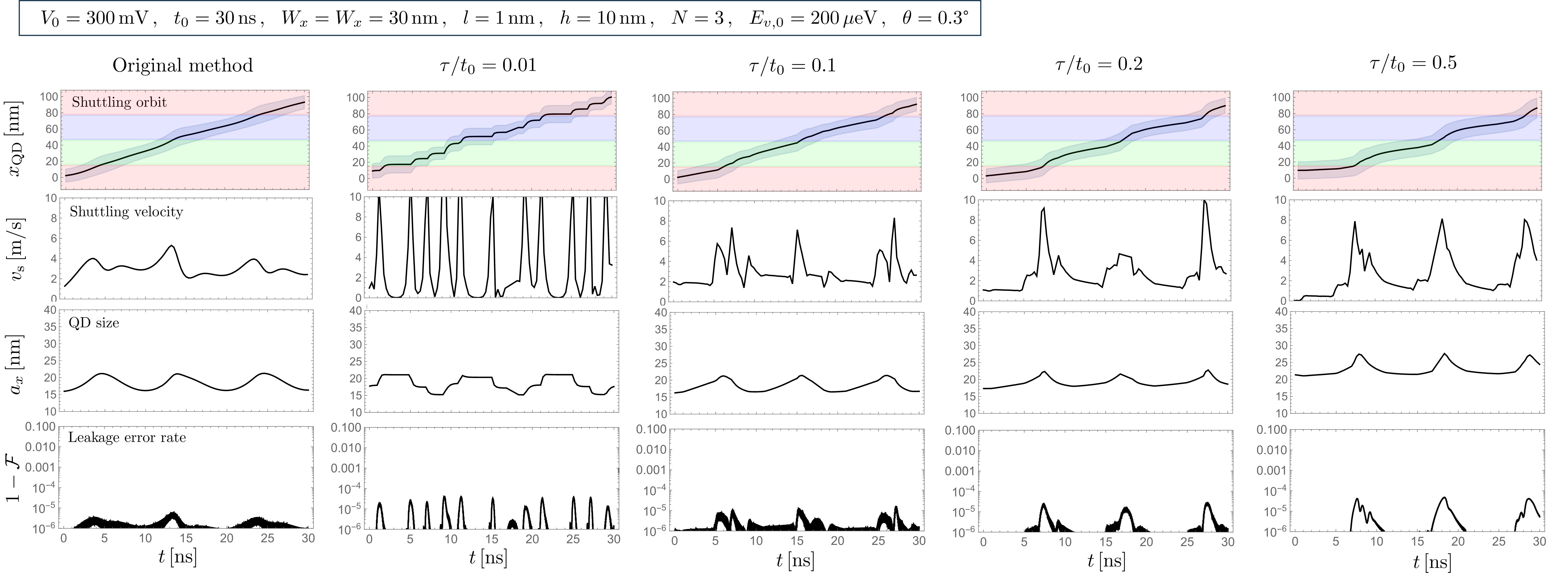}
  \caption{Numerical results comparing shuttling performance for various $\tau$ values to the original (sinusoidal) method. We fix $t_0=30\,\mathrm{ns}$, $W=30\,\mathrm{nm}$, $l=1\,\mathrm{nm}$, $N=3$, and consider $V_0=200\,\mathrm{mV}$ (top panel) and $V_0=300\,\mathrm{mV}$ (bottom panel). The leftmost columns are the original sinusoidal method, while the others are the proposed digital approach with $\tau/t_0=0.01,0.1,0.2,0.5$. For each case, we show $x_{\rm{QD}}$, $v_{\mathrm s}$, $a_x$, and $1-\mathcal{F}$.}
  \label{fig-sim-tau}
\end{figure*}

Figure~\ref{fig-sim-tau} shows how the time constant $\tau$ affects the shuttling fidelity. Each set of panels shows $x_{\rm{QD}}$, $v_{\mathrm s}$, $a_x$, and $1-\mathcal{F}$ for $\tau/t_0=0.01,0.1,0.2,0.5$, compared with the conventional method. Here we employ the same model of valley coupling with figure \ref{fig-sim-proposed}. We fix $t_0=30\,\mathrm{ns}$, $W=30\,\mathrm{nm}$, $l=1\,\mathrm{nm}$, and $N=3$. We set $V_0=200\,\mathrm{mV}$ (top) and $V_0=300\,\mathrm{mV}$ (bottom). We find that the proposed method achieve the high-fidelity comparable the conventional method when $\tau/t_0\simeq0.1$.
For small $\tau/t_0$, the waveform becomes non-smooth resulting the loss of fidelity.
For large $\tau/t_0$, the voltage amplitude is effectively reduced by the long RC time, leading to larger $a_x$ and slight degradation in fidelity. Comparing $V_0=200\,\mathrm{mV}$ to $V_0=300\,\mathrm{mV}$ also confirms that shuttling orbit ($x_{\rm{QD}}, v_{\rm{s}}$) is unchanged while the quantum dot size $a_x$ is smaller for larger $V_0$, consistent with the discussion in Sec.~\ref{sec:CVshuttling}.

\begin{figure*}[t]
  \centering
  \includegraphics[width=18cm]{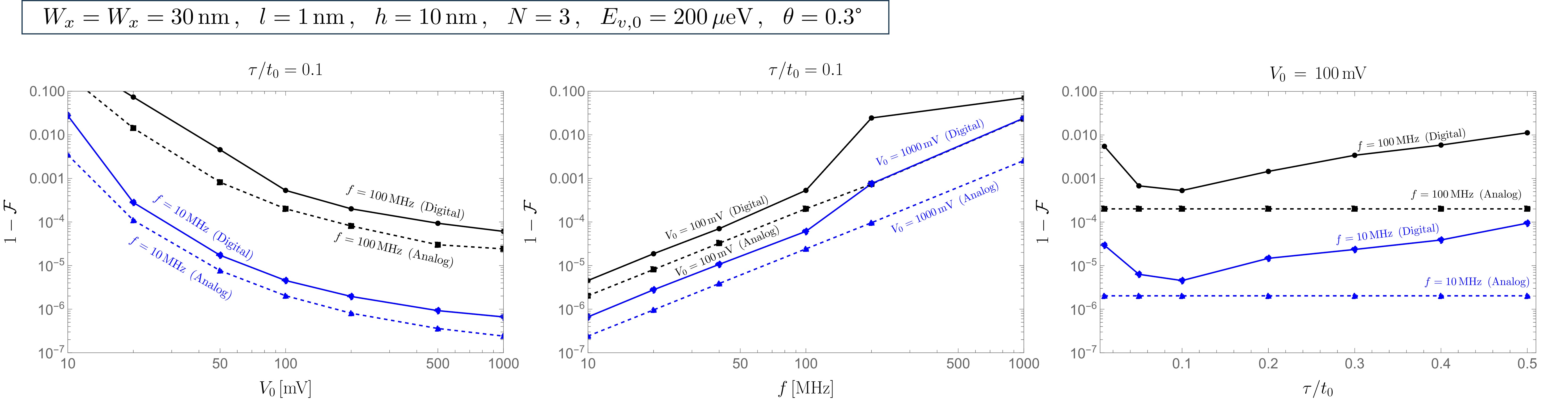}
  \caption{Comparison of shuttling infidelity $(1-\mathcal{F})$ between the conventional (“Analog”) and proposed (“Digital”) methods. We show dependence on $V_0$ (left), $f$ (center), and $\tau$ (right). In the left panel, black lines are $f=100\,\mathrm{MHz}$, blue lines $10\,\mathrm{MHz}$. In the center panel, black lines are $V_0=100\,\mathrm{mV}$, blue lines $1\,\mathrm{V}$. We set $W=30\,\mathrm{nm}$, $l=1\,\mathrm{nm}$, $h=10\,\mbox{nm}$, $N=3$, $E_{v,0}=200\,\mu\mathrm{eV}$, $\theta=0.3^\circ$.}
  \label{fig-F-param}
\end{figure*}

Finally, we compare the parameter dependence of the leakage error rate in the conventional and proposed method. 
Fig.~\ref{fig-F-param} shows the maximum leakage rate $(1-\mathcal{F})$ for both methods, as a function of $V_0$ (left), $f$ (center), and $\tau$ (right). We fix $W=30\,\mathrm{nm},\,l=1\,\mathrm{nm},\,N=3$, and $(E_{v,0},\theta)=(200\,\mu\mathrm{eV},\,0.3^\circ)$. 
The solid lines are the proposed approach, the dashed lines the conventional approach. The left panel compares $f=10\,\mathrm{MHz}$ (blue) and $f=100\,\mathrm{MHz}$ (black), while the center panel compares $V_0=100\,\mathrm{mV}$ (black) and $V_0=1\,\mathrm{V}$ (blue).

In all cases, the dependence on $V_0$ and $f$ is qualitatively the same for both methods, 
indicating that the leakage error is primarily governed by the quantum dot size and shuttling velocity.
Meanwhile, the proposed method’s dependence on $\tau$ is relatively mild, 
demonstrating its robustness to practical variations in RC filter parameters, such as time constant spreads. Further circuit simulations performed in Appendix~\ref{app:circuitsim} confirm that typical parameter spreads do not degrade fidelity significantly. Overall, our results confirm that the digital approach can reliably maintain $\mathcal{F}\gtrsim 99.9\%$ under realistic conditions. This makes it a viable and scalable alternative to the conventional sinusoidal scheme.

%------------------------------------------------------------------
\subsection{Heat dissipation}
\label{sec:proposed-power}

To avoid overwhelming the dilution refrigerator’s cooling capacity, the total power dissipated in the sub-$100\,m$K stage must typically be kept below a few $m$W \cite{2017-Vandersypen-et-al}. Here, we estimate the heat dissipation of placing the SWMX and LPF at cryogenic temperatures ({\it{e.g.,}} 100\,$m$K) and show that it can be maintained below sub-$m$W for millions of qubits, under realistic assumptions.

The primary source of heat dissipation originates from the resistor $R$ 
in series with the capacitor ${C}$~\cite{2022-Boter-et-al}. Consider a single cycle of the AC signal, 
split into two phases: charging and discharging the capacitor. 
During the charging half-cycle, the current is 
$i(t) = (V_0/R) \,e^{-t/(R {C})},
$
and the instantaneous power is $P(t) = i^2(t)\,R$. 
Here, $V_0$ denotes the amplitude of the applied gate voltage. 
By integrating $P(t)$ over the charging half-cycle, 
one finds that the energy dissipated is $\tfrac{1}{2}{C}V_0^2$. 
During the discharging half-cycle, another $\tfrac{1}{2}{C}V_0^2$ 
is lost, making the total energy consumption per full cycle $CV_0^2$. 

This analysis shows that each time the capacitor is charged and discharged 
through $R$, $CV_0^2$ of energy is converted into heat. 
In the context of conveyor-belt shuttling, this process repeats at frequency $f$, 
so the average power consumption per capacitor is $CV_0^2 f$. 
In our analysis, we consider three relevant capacitances: 
(1) stray parasitic capacitance in the wiring 
from the DC feedthrough to the SWMX (${C}_{\mathrm{SC}}$), (2) the gate capacitance of the qubit device itself (${C}_{\mathrm{gate}}$), and 
(3) the internal capacitance associated with 
the switch matrix and filters (${C}_{\mathrm{SL}}$).

We first estimate the heat dissipation from the stray parasitic capacitance. We consider through-silicon via (TSV) connections and assume that its typical capacitance is given as ${C}_{\mathrm{SC}} = 10\,\mathrm{fF}$ per single connection. 
From Fig.~\ref{fig-structure}, the system might have $N_{\rm{qubit}}N$ connections at sub-$100\,m$K. 
Then the power dissipation $\mbox{P}_{\mathrm{SC}}$ from parasitic capacitance is estimated as
\begin{widetext}
\begin{align}
\mbox{P}_{\mathrm{SC}}
&\,=\,
N_{\rm{qubit}}N
{C}_{\mathrm{SC}}
V_0^2 f
\nn\\
&\,\simeq\,
3.0
\,\mathrm{mW}
\left(\frac{N_{\rm{qubit}}}{10^6}\right)
\left(\frac{N}{3}\right)
\left(\frac{{C}_{\mathrm{SC}}}{10\,\mathrm{fF}}\right)
\left(\frac{V_0}{10\,\mathrm{mV}}\right)^2
\left(\frac{f}{100\,\mathrm{MHz}}\right).
\label{eq:PSC}
\end{align}
\end{widetext}
Here we set the typical voltage and frequency 
as $V_0=10\,\mbox{mV}$ and $f=100\,\mbox{MHz}$, which can achieve $\mathcal{F}=99.9\%$ fidelity with $v_{\rm{s}}\simeq 10\,\mbox{m/s}$ in a $W=20\,\mbox{nm}$ device as seen in figure \ref{fig-random-valley}. 
We find that, even for a system with $N_{\rm{qubit}}=10^6$, the power dissipation remains a few $m$W, which is well within the cooling capacity of typical dilution refrigerators.

We next focus on the heat dissipation from the gate capacitances.
We assume that the typical capacitance is of order ${C}_{\mathrm{gate}} = 1\,\mathrm{fF}$.
Then, we find
\begin{widetext}
\begin{align}
\mbox{P}_{\mathrm{gate}}
&\,=\,
N_{\rm{qubit}}\,N_{\mathrm{gate}}
{C}_{\mathrm{gate}}
V_0^2 f
\nn\\
&\,\simeq\,
4.0\,\mathrm{mW}
\left(\frac{N_{\rm{qubit}}}{10^6}\right)
\left(\frac{N_{\mathrm{gate}}}{40}\right)
\left(\frac{{C}_{\mathrm{gate}}}{1\,\mathrm{fF}}\right)
\left(\frac{V_0}{10\,\mathrm{mV}}\right)^2
\left(\frac{f}{100\,\mathrm{MHz}}\right)\,,
\end{align}
\end{widetext}
Again, we find that the heat dissipation for $N_{\rm{qubit}}=10^6$ is within the cooling capacity of typical dilution refrigerators.

Lastly, we estimate the internal capacitance of the switches and filters. 
Considering the appropriate time constant $\tau$ is order $\tau/t_0\simeq 0.1$ for achieving high-fidelity, 
the typical capacitance is estimated as ${C}_{\mathrm{SL}}=1\,\mathrm{fF}$ per switch transistor. Thus, for $N_{\mathrm{DC}}=3$, the heat dissipation leads to
\begin{widetext}
\begin{align}
\mbox{P}_{\mathrm{SL}}
&\,=\,
N_{\rm{qubit}}\,N\,N_{\mathrm{DC}}
{C}_{\mathrm{SL}}
V_0^2 f
\nn\\
&\,\simeq\,
0.9\,\mathrm{mW}
\left(\frac{N_{\rm{qubit}}}{10^6}\right)
\left(\frac{N}{3}\right)
\left(\frac{N_{\mathrm{DC}}}{3}\right)
\left(\frac{{C}_{\mathrm{SL}}}{1\,\mathrm{fF}}\right)
\left(\frac{V_0}{10\,\mathrm{mV}}\right)^2
\left(\frac{f}{100\,\mathrm{MHz}}\right).
\label{eq:PSL}
\end{align}
\end{widetext}
We observe that the heat dissipation for $N_{\rm{qubit}}=10^6$ is within the cooling capacity of typical dilution refrigerators.

\begin{figure}[t]
  \centering
  \includegraphics[width=8cm]{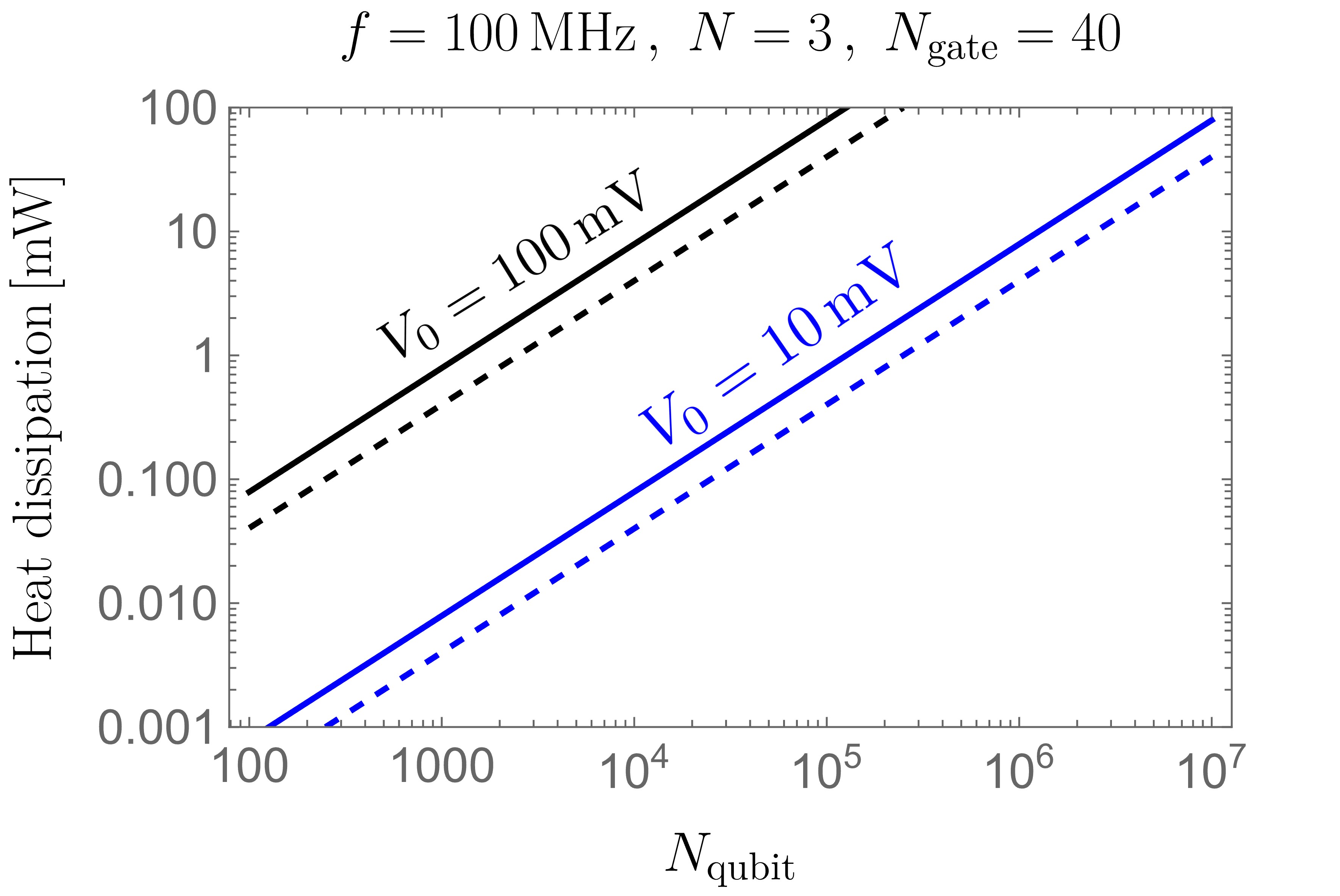}
  \caption{Total heat dissipation as a function of the number of qubit. Here we fix $N=3\,,N_{\rm{gate}}=40\,,C_{\rm{SC}}=10\mbox{fF}\,,C_{\rm{gate}}=C_{\rm{SL}}=1\mbox{fF}\,,$ and $f=100\,\mbox{MHz}$. The black (blue) line corresponds to the case with $V_0\,=\,10\,(100)\,\mbox{mV}$.
  The dotted lines correspond to the results where we set $N_{\rm{qubit}}=1$ in Eqs.~(\ref{eq:PSC}) and (\ref{eq:PSL}).}
  \label{fig-heat}
\end{figure}
Combining these estimates under typical conditions, {\it{e.g.}}, $V_0=10\,\mathrm{mV}$ and $f=100\,\mathrm{MHz}$ (achieving $v_{\mathrm s}\sim 00\,\mathrm{m/s}$ and $\mathcal{F}\sim 99\%$ in $W=20\,\mbox{nm}$ [Fig.~\ref{fig-random-valley}]), shows that the total sub-mW heat load is feasible for up to $N_{\rm{qubit}}\simeq 10^6$. 
Figure \ref{fig-heat} shows the estimated total heat dissipation ($=\mbox{P}_{\rm{SC}}+\mbox{P}_{\rm{gate}}+\mbox{P}_{\rm{SL}}$) as a function of the number of qubit{\footnote{
We note that Eqs. (\ref{eq:PSC}) and (\ref{eq:PSL}) depend on the specific architecture of the Si device. For example, in the architecture proposed in Ref. \cite{2024-SpinBus}, multiple shuttling channels can be driven using shared control lines, which would eliminate the linear dependence on the number of qubits in these equations.
In our current estimates, we adopt a conservative approach to avoid underestimating the heat dissipation, assuming that each qubit requires its own control line. The dotted lines in figure \ref{fig-heat} correspond to the results where we set $N_{\rm{qubit}}=1$ in Eqs.~(\ref{eq:PSC}) and (\ref{eq:PSL}).
}}. Here we fix $N=3\,,N_{\rm{gate}}=40\,,C_{\rm{SC}}=10\mbox{fF}\,,C_{\rm{gate}}=C_{\rm{SL}}=1\mbox{fF}\,,$ and $f=100\,\mbox{MHz}$. The black (blue) line corresponds to the case with $V_0\,=\,10\,(100)\,\mbox{mV}$ achieving $v_{\mathrm s}\sim 10\,\mathrm{m/s}$ and $\mathcal{F}\sim 99\%$ in $W=20\,(30)\,\mbox{nm}$ case.

The number of qubits can potentially be increased further through additional optimizations. One  promising approach is to reduce $V_0$. As evaluated above, the power dissipation for each qubit is proportional to $N_{\rm{qubit}} V_0^2$. Therefore, reducing $V_0$ allows for an increase in the permissible number of qubits. Maintaining high fidelity while reducing $V_0$ can be achieved by decreasing the gate size, {\it{i.e.}}, pursuing further miniaturization of the device.
Another consideration is that the above estimation assumes individual control of each qubit (or quantum dot). If collective control of $N_{\rm{unit}}$ ($<N_{\rm{qubit}}$) qubits per unit is allowed, the power dissipation formula can be corrected as $N_{\rm{qubit}} \to N_{\rm{qubit}} / N_{\rm{unit}}$. In this case, the permissible number of qubits can also be increased.

%%%%%%%%%%%%%%%%%%%%%%%%%%%%%%%%%%%%%%%%%%%%%%%%%%%%%%%%%%%%%%%%%%%
\section{Conclusion and Discussion}
\label{sec:conclusion}
We have proposed a digital-controlled conveyor-belt shuttling approach for silicon spin qubits that addresses the major scaling challenges of the conventional analog sinusoidal method. By placing a switch matrix (SWMX) and low-pass filter (LPF) in a cryogenic environment, we can synthesize approximate sinusoidal waveforms from a small set of DC voltage levels. Our numerical simulations indicate that this method achieves shuttling fidelities of $>99.9\%$, which is compatible with those of the conventional analog approach.

Key advantages of this proposal include:
\begin{itemize}
\item \textbf{Significant reduction in wiring complexity.} 
%In the conventional scheme, each gate electrode typically requires a high-frequency line from room temperature, leading to $\mathcal{O}(N_{\rm{qubit}})$ cables if the device inhomogeneities present. 
Our proposal relies on only a few DC lines plus a few control lines at room temperature, effectively avoiding the wiring crowding.

\item \textbf{Simplified compensation of high-frequency distortions.}
Waveform generation at cryogenic temperature obviates substantial frequency-dependent losses and phase shifts in the signal lines, thereby reducing the need for complex calibrations.

\item \textbf{High fidelity and robustness.} 
We demonstrated that even with approximate rectangular pulses and simple RC filtering, the leakage error rate remains below $10^{-4}$, offering strong resilience to device variations in RC time constants and other parameters.

\item \textbf{Below-sub\,mW heat dissipation in the sub-100\,mK stage.} 
Our estimates based on capacitor charging suggest total power dissipation can remain under sub-\,mW for millions of qubits under realistic conditions, compatible with the cooling budget of a dilution refrigerator.
\end{itemize}

These results highlight the promise of digital-controlled conveyor-belt shuttling for integrating large numbers of silicon spin qubits while maintaining high coherence. Future directions include:

\begin{itemize}
\item \textbf{Flexibility in shuttling operation.}
Digital control offers fine-grained switching of voltage phases. This could be advantageous for flexible shuttling (including the changing the shuttling velocity and/or direction \cite{2024-SpinBus,2024-Losert-et-al,2024-Nemeth-et-al}) which may be important for advanced QEC protocols.

\item \textbf{Experimental demonstration and further device miniaturization.}
While our numerical analysis is encouraging, experimental validation is crucial. Reducing gate size can yield smaller quantum dots at lower voltages, suppressing valley leakage and heat dissipation.

\item \textbf{Circuit-level optimizations.}
Although we used a simple capacitor-based power estimate, further engineering of SWMX and LPF designs could reduce power consumption and noise even more.

\end{itemize}

Overall, the proposed digital-control framework paves the way for scalable, high-fidelity spin shuttling in silicon. By combining CMOS-compatible fabrication with cryogenic electronics, this approach strengthens the foundation for fault-tolerant quantum computation in large-scale silicon spin qubit architectures.

\begin{acknowledgments}
We thank Shinobu Yabuki, Satoru Akiyama, and Tatsuya Tomaru
for helpful discussions.
We also thank Jeremy Garaffa for pointing out a typo in the first version of our manuscript.
This work is supported by JST Mooshot R\&D Grant No. JPMJMS2065.
\end{acknowledgments}

%%%%%%%%%%%%%%%%%%%%%%%%%%%%%%%%%%%%%%%%%%%%%%%%%%%%%%%%%%%%%%%%%%%
\appendix
\section{A moving quantum dot and gate voltages}
\label{app:VQD}
Let us consider a quantum device shown in the top left of Fig.~\ref{fig-sim-proposed}.
Each gate electrode has a common size of 
\(W_x \times W_y\) and that the spacing between the silicon layer $l$ and the gate electrode 
layer is \(h\), then the Coulomb potential \(V_{\rm{QD}}(x,y)\) formed in the silicon layer 
(in the \((x,y)\) plane) can be estimated as
\begin{widetext}
\begin{align}
V_{\rm{QD}}(x,y)
\,=\,
-e
\sum_{i=1}^{N_{\rm{gate}}}\,
\frac{\rho_i}{4\pi\epsilon_{\rm{Si}}}\,
\int_{-\frac{W_y}{2}}^{\frac{W_y}{2}} dY
\int_{X_{Li}}^{X_{Ri}} dX\,
\frac{1}{\sqrt{(x-X)^2+(y-Y)^2+h^2}}
\,,
\end{align}
\end{widetext}
with
\begin{align}
X_{Li}
\,=\,
-\frac{W_x}{2}+W_x\,l\,(i-1),\quad
X_{Ri}
\,=\,
\frac{W_x}{2}+W_x\,l\,(i-1).
\label{eq:V}
\end{align}
Here, \(\epsilon_{\rm{Si}}\) is the permittivity of silicon, given by 
\(\epsilon_{\rm{Si}}\simeq 11.68\,\epsilon_0\), where \(\epsilon_0\) is the vacuum permittivity. 
$e$ is the electron charge.
\(N_{\rm{gate}}\) denotes the total number of gate electrodes, and \(\rho_i\) is the charge density 
on the surface of the \(i\)th gate electrode. For simplicity, we assume that the charge is 
uniformly distributed over the electrode surface, so that $
\rho_i = {C_i V_i}/{S_i}$ with $ S_i = W_xW_y$.
Here, \(V_i\) is the voltage applied to gate electrode \(i\), and \(C_i\) is the effective capacitance 
of gate electrode \(i\). In the subsequent analysis, we assume that \(C_i\) is identical for all 
gates, and we take its typical value to be \(0.1\,\mathrm{aF}\).

By appropriately choosing the gate voltages \((V_1,\cdots,V_{N_{\rm{tot}}})\), the potential 
\(V_{\rm{QD}}(x,y)\) can be engineered to exhibit multiple minima, with each minimum corresponding to a quantum dot \cite{1995-Davies-et-al}.

Let us focus on a single quantum dot.
Ignoring quantum correlations with other quantum dots, 
the quantum dot typically takes a harmonic-oscillator-like shape, approximated by
\begin{align}
V_{\rm{QD}}(x,y)
\,\simeq\,
\frac{m_x\omega_x^2}{2}(x-x_{\rm{QD}})^2
+
\frac{m_y\omega_y^2}{2}(y-y_{\rm{QD}})^2,
\label{eq:VQD1}
\end{align}
where $m_x$ and $m_y$ are the effective masses of electrons in the silicon layer, typically $m_x = m_y \approx 0.19\,m_e$ (with $m_e\simeq 0.5\,\mathrm{MeV}$ the electron mass in vacuum). 
We interpret the parameters $(x_{\rm{QD}},y_{\rm{QD}},\omega_x,\omega_y)$ as the quantum dot’s position and $(\omega_x,\omega_y)$ representing energy scales of the quantum dot. 
The quantum dot sizes $a_x$ and $a_y$ are related to $\omega_x$ and $\omega_y$ by
\begin{align}
a_x = \sqrt{\frac{1}{m_x\omega_x}},\quad
a_y = \sqrt{\frac{1}{m_y\omega_y}}.
\end{align}

The parameters $(x_{\rm{QD}},y_{\rm{QD}},\omega_x,\omega_y)$ depend on the applied gate voltages.
The position of a quantum dot, \((x_{\rm{QD}}(t),y_{\rm{QD}}(t))\), 
is identified with a minimum of \(V_{\rm{QD}}(x,y)\) [Eq.~(\ref{eq:V})], and the quantum dot sizes \((a_x(t),a_y(t))\) 
are estimated via
\begin{widetext}
\begin{align}
a_x(t)
\,=\,
\sqrt{\frac{1}{m_x\,\omega_x(t)}}, \qquad
\omega^2_x(t)
\,=\,
\frac{1}{m_x}
\frac{\partial ^2V_{\rm{QD}}}{\partial x^2}\Biggl|_{x=x_{\rm{QD}}(t),\,y=y_{\rm{QD}}(t)}
\,,\\
a_y(t)
\,=\,
\sqrt{\frac{1}{m_y\,\omega_y(t)}}, \qquad
\omega^2_y(t)
\,=\,
\frac{1}{m_y}
\frac{\partial ^2V_{\rm{QD}}}{\partial y^2}\Biggl|_{x=x_{\rm{QD}}(t),\,y=y_{\rm{QD}}(t)}.
\end{align}
\end{widetext}

Shuttling is realized by time-varying the gate voltages, thereby modifying the 
quantum dot shape in the silicon layer. In particular, when only the parameters 
$\bigl(x_{\rm{QD}}, y_{\rm{QD}}, \omega_x, \omega_y\bigr)$ in Eq.~\eqref{eq:VQD1} 
are varied over time while retaining the dot’s fundamental harmonic form, 
we refer to this process as \emph{conveyor-belt shuttling}~\conveyorall\footnote{Another approach, exploiting tunneling between adjacent dots, is sometimes called \emph{bucket-brigade} shuttling~\bucketall.}.

%%%%%%%%%%%%%%%%%%%%%%%%%%%%%%%%%%%%%%%%%%%%%%%%%%%%%%%%%%%%%%%%%%%
\section{Fidelity of conveyor-Belt Spin Shuttling in Silicon}
\label{sec:CVshuttling}
\textbf{}

Let us next estimate shuttling fidelity.
Although various mechanisms can alter the electron's quantum state during shuttling, 
we focus here on transitions into valley-excited states, 
which are known to be the dominant factor limiting the maximum shuttling speed~\cite{2023-Langrock-et-al,2024-Losert-et-al,2024-David-et-al,2024-Oda-et-al,2024-Nemeth-et-al}. 
In silicon-based spin qubit architectures, 
the electron valley degree of freedom should ideally remain in its ground state 
$\ket{v_g}$; any leakage into the excited valley state $\ket{v_e}$ reduces quantum computational fidelity. 
Below, we evaluate the leakage error rate caused by shuttling and discuss strategies to suppress it.

Let $\ket{\Psi_v(t)} = \alpha_g(t)\ket{v_g} + \alpha_e(t)\ket{v_e}$ be the valley state at time $t$ and introduce
\begin{align}
\ket{\Psi_v(t)} 
\,=\,
\begin{bmatrix}
\alpha_g(t) \\
\alpha_e(t)
\end{bmatrix}
\,.
\end{align}
The time evolution of the coefficients $(\alpha_g, \alpha_e)$ follows the time-dependent Schr\"odinger equation:
\begin{align}
i \frac{d}{dt}
\begin{bmatrix}
\alpha_g(t) \\
\alpha_e(t)
\end{bmatrix}
=
\frac{1}{2}
\begin{bmatrix}
 -E_v(t) & -\dot{\phi}_v(t) \\
 -\dot{\phi}_v(t) & E_v(t)
\end{bmatrix}
\begin{bmatrix}
\alpha_g(t) \\
\alpha_e(t)
\end{bmatrix},
\label{eq:Scheq-valley}
\end{align}
where $E_v$ and $\phi_v$ are defined from the inter-valley coupling $\Delta(t)$ via
\begin{align}
E_v(t) = 2|\Delta(t)|,\quad
\phi_v(t) = \mathrm{Arg}\,\Delta(t).
\end{align}
%(See Appendix~\ref{app:Hv} for derivation.)

The shuttling leakage error rate $(1 - \mathcal{F})$ can be evaluated by solving 
Eq.~(\ref{eq:Scheq-valley}). We assume the initial condition 
$\ket{\Psi_v(0)} = \ket{v_g}$ and determine the time-evolved state 
$\ket{\Psi_v(t)}$ from Eq.~(\ref{eq:Scheq-valley}). In this way, the fidelity is given by 
$\mathcal{F} = \bigl|\braket{v_g| \Psi_v(t)}\bigr|^2$, 
so $(1 - \mathcal{F})$ corresponds to the probability of leakage into excited 
valley states. Recent studies suggest that achieving 
$\mathcal{F} \gtrsim 99.9\%$ is necessary for implementing quantum error correction 
in realistic silicon architectures~\cite{2024-Siegel-et-al}.

The valley coupling $\Delta$ depends on both the silicon interface and the shape of the quantum dot. Theoretically, it is given by~\cite{2006-Friesen-Eriksson-Coppersmith,2007-Friesen-Chutia-Tahan-Coppersmith,2006-Nestoklon-Golub-Ivchenko,2011-Saraiva-et-al,2022-Wuetz-Losert-Coppersmith-Friesen-et-al,2023-Losert-Coppersmith-Friesen-et-al}
\begin{align}
\Delta\,=\,
\int d^3\vec{r}\, e^{-2ik_0z}\, |\psi_{\rm{QD}}(\vec{r})|^2\, V_{\rm{QW}}(\vec{r})
\,.
\label{eq:delta}
\end{align}
Here $k_0$ characterizes the wavevector of the valley states. It is related to the Si lattice spacing $a_0=0.53\,\mathrm{nm}$ via 
\begin{align}
k_0 = 0.85\,\frac{2\pi}{a_0} \,\simeq\, (0.1\,\mathrm{nm})^{-1}.
\end{align}
$\psi_{\rm{QD}}(\vec{r})$ represents the wavefunction of an electron confined in the quantum dot, and it is assumed that the spatial extent of this wavefunction is much larger than $1/k_0 \simeq 0.1\,\mathrm{nm}$. The function $V_{\rm{QW}}(\vec{r})$ denotes the confinement potential along the [001] (i.e., $z$) crystallographic direction.

The integrand in Eq.~(\ref{eq:delta}) contains the oscillatory term $e^{-2ik_0z}$, which arises from the interference of the Bloch functions corresponding to the valley states. Since this oscillatory term varies on a length scale of approximately $1/k_0 \simeq 0.1\,\mathrm{nm}$, contributions from those parts of $V_{\rm{QW}}(\vec{r})$ that are smooth on the $1/k_0$ scale cancel out. Therefore, only the structural details of $V_{\rm{QW}}(\vec{r})$ in the vicinity of the silicon interface contribute to $\Delta$. This indicates that $\Delta$ is determined by both the interface structure of silicon and the shape of the quantum dot.

%%%%%%%%%%%%%%%%%%%%%%%%%%%%%%%%%%%%%%%%%%%%%%%%%%%%%%%%%%%%%%%%%%%
\subsection{Smooth and tilted interface model}
\label{app:delta}
In this section, we calculate $\Delta$ under the assumption that the silicon interface is a flat plane tilted by an angle $\theta$ in the shuttling direction (the $x$-axis). 
In this case, the interface is described by
\[
z=z_0(x)=z_0 - x\,\tan\theta \simeq z_0 - x\theta.
\]
Substituting 
\[
V_{\rm{QW}}(\vec{r}) = v\,\delta(z-z_0(x))
\]
into Eq.~(\ref{eq:delta}) and performing the integrations over $y$ and $z$, we obtain
\begin{align}
\Delta
\,=\,
v\int dx\, e^{-2ik_0(z_0-x\theta)}\, |\psi_{x}(x)|^2\, |\psi_{z}(z_0-x\theta)|^2
\,.
\end{align}
Here, we have assumed that the quantum dot wavefunction factorizes as
\[
\psi_{\rm{QD}}(\vec{r})=\psi_x(x)\,\psi_y(y)\,\psi_z(z).
\]
Furthermore, if we approximate 
\[
|\psi_{z}(z_0-x\theta)|^2 \simeq |\psi_{z}(z_0)|^2,
\]
then
\begin{align}
\Delta
\,=\,
\Delta_0\,\int dx\, e^{2ik_0x\theta}\, |\psi_{x}(x)|^2
\,,
\end{align}
where $\Delta_0$ is defined by
\[
\Delta_0 = v\,|\psi_{z}(z_0)|^2\,e^{-2ik_0z_0},
\]
which corresponds to the valley coupling when $\theta=0$.

Finally, taking
\begin{align}
|\psi_x(x)|^2\,=\,
\frac{1}{\sqrt{\pi}\,a_x}\exp\Bigl(-\frac{(x-x_{\rm{QD}})^2}{a_x^2}\Bigr),
\end{align}
and performing the integration over $x$, we obtain
\begin{align}
\Delta
\,=\,
\Delta_0\, e^{-k_0^2\theta^2 a_x^2}\, e^{2ik_0\theta x_{\rm{QD}}}.
\end{align}
Thus, Eqs.~(\ref{eq:Ev}) and (\ref{eq:phiv}) are derived. Note that this result corresponds to the linear gradient model (or smoothly tilted interface model \cite{2007-Friesen-Chutia-Tahan-Coppersmith}) as described in Ref.~\cite{2023-Langrock-et-al}.

Under these assumptions, $E_v$ and $\dot{\phi}_v$ can be written in terms of the quantum dot size $a_x$ and the shuttling velocity $v_{\rm s}(t)=\dot{x}_{\rm{QD}}(t)$ as~\cite{2023-Langrock-et-al}:
\begin{align}
E_v(t) 
&= 
E_{v,0}\exp\left[-k_0^2\theta^2a_x^2(t)\right],
\label{eq:Ev}
\\
\dot{\phi}_v(t)
&=
2k_0 \theta v_{\rm s}(t),
\label{eq:phiv}
\end{align}
$E_{v,0}$ is valley splitting with $\theta=0$, typically on the order of $\mathcal{O}(100\,\mu\mathrm{eV})$~\cite{2013-Zwanenburg-et-al,2023-Burkard-et-al,2024-Volmer-et-al}.
Unless otherwise specified, we set $E_{v,0}=200\,\mu$eV in according to Ref.~\cite{2023-Langrock-et-al}.

We next identify the conditions necessary to achieve $\mathcal{F}\gtrsim 99.9\%$. From Eq.~(\ref{eq:Scheq-valley}), we see that one can suppress unwanted excitations by making the energy gap $E_v$ large and/or the effective driving term $\dot{\phi}_v$ small. From Eqs.~(\ref{eq:Ev}) and (\ref{eq:phiv}), this can be achieved by decreasing the quantum dot size $a_x$ or slowing the shuttling velocity $v_{\mathrm s}$.
We now provide a concrete example to clarify these conditions quantitatively. 
Suppose that $a_x$ and $v_{\mathrm s}$ remain constant in time. 
Under this assumption, Eq.~(\ref{eq:Scheq-valley}) reduces to a two-level 
Rabi formula whose transition probability can be estimated analytically.
The leakage error (maximum transition probability) is thus given as~\cite{2023-Langrock-et-al}
\begin{align}
1-\mathcal{F}
=
\frac{4k_0^2\theta^2v_{\mathrm s}^2}{4k_0^2\theta^2v_{\mathrm s}^2 + E_{v,0}^2\exp[-2k_0^2\theta^2 a_x^2]}
\,.
\label{eq:F-eq}
\end{align}
We note that the leakage error rate exhibits an exponential dependence on $a_x$. 
This exponential sensitivity implies that achieving high fidelity requires 
precise control over the quantum dot size.

In summary, we have introduced the general aspects of conveyor-belt shuttling in silicon 
and discussed the conditions required for high-fidelity operation based on smooth and tilted interface model. 
We emphasized that the shuttling fidelity is particularly sensitive 
to the quantum dot size $a_x$, which implies the need for precise control of $a_x$ 
to achieve fidelities above 99.9\%. 
These findings guide the choice of operating parameters necessary 
for realizing high-fidelity conveyor-belt shuttling.

Before closing this section, let us comment on the model of valley coupling employed here. 
While the smooth and tilted silicon interface model simplifies the actual interface, it provides valuable insights into the fundamental relationship between shuttling fidelity, quantum dot size, and shuttling velocity, such as Eqs.~(\ref{eq:Ev}), (\ref{eq:phiv}), and (\ref{eq:F-eq}).  
To obtain a more realistic assessment, it is necessary to incorporate the effects of interface roughness, which would inevitably complicate the model.  
Fidelity calculations considering interface roughness have been performed in the Appendix~\ref{app:delta-random}, demonstrating that the proposed digital-controlled shuttling method maintains high fidelity even under these more realistic conditions.

%%%%%%%%%%%%%%%%%%%%%%%%%%%%%%%%%%%%%%%%%%%%%%%%%%%%%%%%%%%%%%%%%%%
\subsection{Interface roughness model} \label{app:delta-random}

\begin{figure*}[t]
  \centering
  \includegraphics[width=17cm]{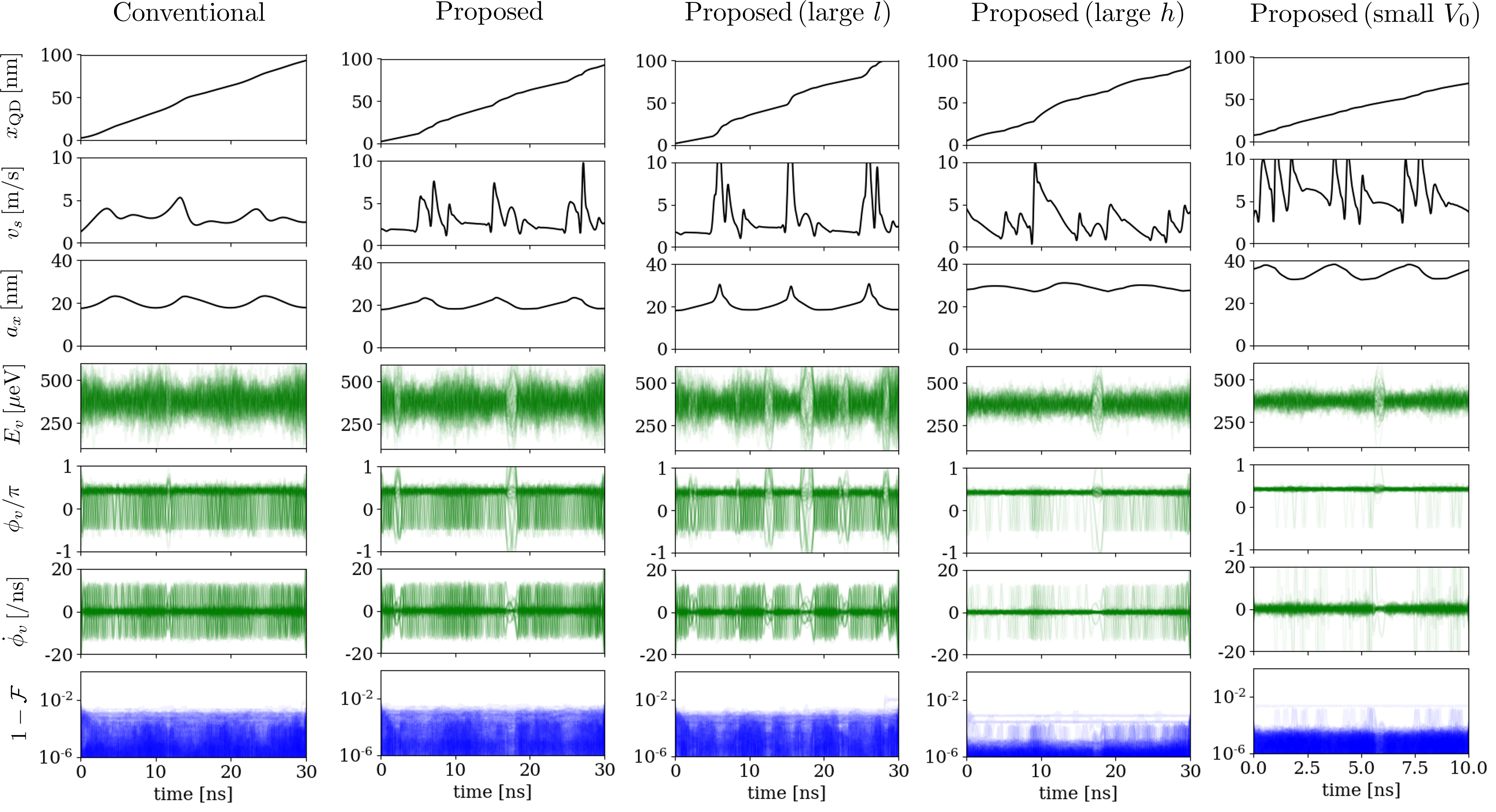}
  \caption{The time evolution of various quantities relevant to the shuttling process. We generated 100 instances of random potential profiles that incorporate interface roughness.  
  The leftmost figure corresponds to the result for the conventional method with \((W, l, h, V_0, t_0) = (30\,\mathrm{nm}, 1\,\mathrm{nm}, 10\,\mathrm{nm}, 200\,\mathrm{mV}, 30\,\mathrm{ns})\).  
  The other figures show results for the proposed method.  
  The parameters from left to right are set as  
  \((W, l, h, V_0, t_0, \tau/t_0) = (30\,\mathrm{nm}, 1\,\mathrm{nm}, 10\,\mathrm{nm}, 200\,\mathrm{mV}, 30\,\mathrm{ns}, 0.1)\),  
  \((30\,\mathrm{nm}, 5\,\mathrm{nm}, 10\,\mathrm{nm}, 200\,\mathrm{mV}, 30\,\mathrm{ns}, 0.1)\),  
  \((30\,\mathrm{nm}, 1\,\mathrm{nm}, 30\,\mathrm{nm}, 200\,\mathrm{mV}, 30\,\mathrm{ns}, 0.1)\), and  
  \((20\,\mathrm{nm}, 1\,\mathrm{nm}, 10\,\mathrm{nm}, 10\,\mathrm{mV}, 100\,\mathrm{ns}, 0.1)\).}
  \label{fig-random-valley}
\end{figure*}

This section evaluates the valley coupling \(\Delta\) in a more realistic setup that includes interface roughness.  
In Si/SiGe heterostructures, alloy disorder causes the interface to become effectively random, leading to fluctuations in the shape of the quantum well potential \(V_{\rm{QW}}\).  
Following the probabilistic models proposed in recent studies \cite{2022-Wuetz-Losert-Coppersmith-Friesen-et-al, 2023-Losert-Coppersmith-Friesen-et-al}, we evaluate \(\Delta\) via numerical simulations that incorporate these interface roughness effects.  
This allows us to discuss how such disorder influences the shuttling fidelity.

The computational procedure is as follows:  
First, we assume that \(V_{\rm{QW}}\) depends only on \(x\) and \(z\), i.e., \(V_{\rm{QW}}(\vec{r})=V_{\rm{QW}}(x,z)\).  
Under this assumption, the wavefunction can be decomposed as
\(\psi_{\rm{QD}}(\vec{r}) = \psi_{x,z}(x,z)\psi_y(y)\).
We also consider that the wavefunction can be effectively decomposed as
$\psi_{x,z}(x,z)\simeq \psi_x(x)\psi_z(z)$
since the mixing between $x$ and $z$ enters only from miscut angle dependence 
along the shuttling direction, $x$.
Hereafter we consider $|\psi_x(x)|^2=(\sqrt{\pi}a_{\rm{QD}})^{-1}\exp(-(x-x_{\rm{QD}})^2/a_{\rm{QD}}^2)$.
We finally have
\begin{align}
\Delta = \int dz\, e^{-2 i k_0 z} |\psi_z(z)|^2 V_{\rm{QW}}(z).
\end{align}  
where 
\begin{align}
V_{\rm{QW}}(z) = \int dx\, |\psi_x(x)|^2 V_{\rm{QW}}(x,z).
\end{align} 
To include interface roughness, 
we discretize the \((x,z)\)-space at the atomic scale, resulting in:  
\begin{align}
\Delta = 
\sum_{l} \delta z\, e^{-2 i k_0 z_l} |\psi_z(z_l)|^2 V_{\rm{QW}}(z_l),
\end{align}  
and
\begin{align}
V_{\rm{QW}}(z_l)
 = 
\sum_{j,l} \delta x\, |\psi_x(x_j)|^2 V_{\rm{QW}}(x_j,z_l),
\end{align}  
where \(\delta z = a_0/4,\,\delta x = a_0/2\).  
The randomness enters through \(V_{\rm{QW}}(x_j,z_l)\).  
Specifically, we model it as:  
\begin{align}
V_{\rm{QW}}(x_j,z_l) = \Delta E_c \frac{X(x_j,z_l) - X_s}{X_s - X_w} + e E_z z_l,
\end{align}  
where \(X(x_j,z_l)\) is a random variable, and \((X_w, X_s, \Delta E_c, E_z)\) are parameters chosen based on recent analyses \cite{2025-Marcks-et-al}.  
In our setup, we set \(X_w=0.0972\), \(X_s=0.3\), \(\Delta E_c=139\,\mathrm{meV}\), and \(E_z=13\,\mathrm{mV/nm}\).  
We define \(X(x_j,z_l)\) as a binomial random variable:  
\begin{align}
X(x_j,z_l) \sim \frac{1}{N_{\rm{eff}}} \text{Binom}(N_{\rm{eff}}, \bar{X}(x_j,z_l)),
\end{align}  
with \(N_{\rm{eff}}=4 \pi a_{\rm{QD}}^2 / a_0^2\), and  
\begin{align}
\bar{X}(x_j,z_l) = X_w 
&+ \frac{X_s - X_w}{1 + \exp[(z_l - z_t(x_j))/\tau]} \nn\\
&+ \frac{X_s - X_w}{1 + \exp[(z_b(x_j) - z_l)/\tau]}.
\end{align}
Here $z_{b,t}(x_j)$ describe the interface and we take
$z_{b,t}(x_j)=z_{b,t}-\theta x_j$ with $\theta$ being miscut angle. 
In our analysis, we set \(z_b - z_t = 80\,\mathrm{ML}\), $\theta=0.3^\circ$ and \(\tau = (1/4)\,\mathrm{ML}\), where \(\mathrm{ML}\) stands for atomic monolayers (\(1\,\mathrm{ML}=a_0/4\)).  
The wavefunction \(\psi_z(z)\) is obtained by numerically solving the Schrödinger equation with the potential $V_{\rm{QW}}(z)$ with replacing \(X(x_j,z_l)\) to $\bar{X}(0,z)$.

The fidelity of the shuttling process is evaluated as follows:  
We generate 100 instances of the random potential profiles using this method, compute \(\Delta\) for each, and simulate the time evolution of the valley state accordingly.  
The fidelity for each run is then calculated, resulting in 100 fidelity values.

Figure \ref{fig-random-valley} shows the results.  
From top to bottom, the panels display: the quantum dot position (\(x_{\rm{QD}}\)), shuttling velocity (\(v_s\)), quantum dot size (\(a_x\)), valley splitting \(E_v\), phase \(\phi_v\), the time derivative of \(\phi_v\), and the shuttling infidelity.  
The leftmost figure shows the results for the conventional method, while the others show the proposed method.  
Specifically, the third panel from the left illustrates the effect of increasing the quantum dot size \(l\), the fourth panel shows the impact of increasing the vertical distance \(h\), and the rightmost panel demonstrates the effect of reducing the control voltage amplitude \(V_0\).

Remarkably, even under this more realistic model, the fidelities of both the conventional and proposed methods remain comparable, with no significant difference.  
This confirms that our main conclusion—that the proposed digital control approach maintains high fidelity even under realistic device conditions—remains valid.

%%%%%%%%%%%%%%%%%%%%%%%%%%%%%%%%%%%%%%%%%%%%%%%%%%%%%%%%%%%%%%%%%%%
\section{Conventional implementation of conveyor-belt shuttling}
\label{sec:conventional}
This section reviews the conventional implementation of conveyor-belt shuttling.
%\begin{figure*}[t]
%  \centering
%  \includegraphics[width=15cm]{fig-V-conventional.jpg}
%  \caption{Voltage waveforms in the conventional (sinusoidal) control for conveyor-belt shuttling. The dotted line shows a basic waveform without phase modulation, while the solid line shows the result with phase modulation.}
%  \label{fig-V-conventional}
%\end{figure*}
In the conventional implementation of conveyor-belt shuttling, 
a phase-modulated sinusoidal voltage is applied to gate electrodes arranged 
along the shuttling direction~\conveyorexp. Concretely, the applied voltage 
on the $i$th gate $V_i(t)$ is given by
\begin{align}
V_i(t) 
= 
V_0 \cos\biggl(2\pi f t + \Delta\phi(t) + \tfrac{2\pi}{N}(i-1)\biggr)\,.
\label{eq:Vsin}
\end{align}
The gate voltages are applied in a periodic 
fashion: specifically, the $(N+1)$-th gate electrode is driven by the same voltage 
as the first gate, $V_1$, the $(N+2)$-th gate by $V_2$, and so forth, creating 
a repeating pattern of length $N$. 
$N$ is often set to 3 or 4.
Here, $V_0$ and $f$ are constant parameters corresponding to the amplitude 
and frequency of the voltage waveform, respectively. As will be shown, $V_0$ 
mainly determines the quantum dot size, while $f$ sets the shuttling velocity. 
The term $\Delta\phi(t)$ provides phase modulation to maintain a nearly uniform 
shuttling speed~\cite{2024-Jeon-Benjamin-Fisher},\footnote{Instead of the phase 
modulation, the bias offsets can also achieve similar effects.} 
and we specifically choose $\Delta\phi(t)=0.07\pi \sin(2N\pi f t)$ 
which yields near-constant-speed shuttling in our simulations. 
%Figure~\ref{fig-V-conventional} illustrates the resulting waveform, where the dotted line 
%corresponds to $\Delta\phi(t)=0$ and the solid line includes phase modulation.

\begin{figure*}[t]
  \centering
  \includegraphics[width=16cm]{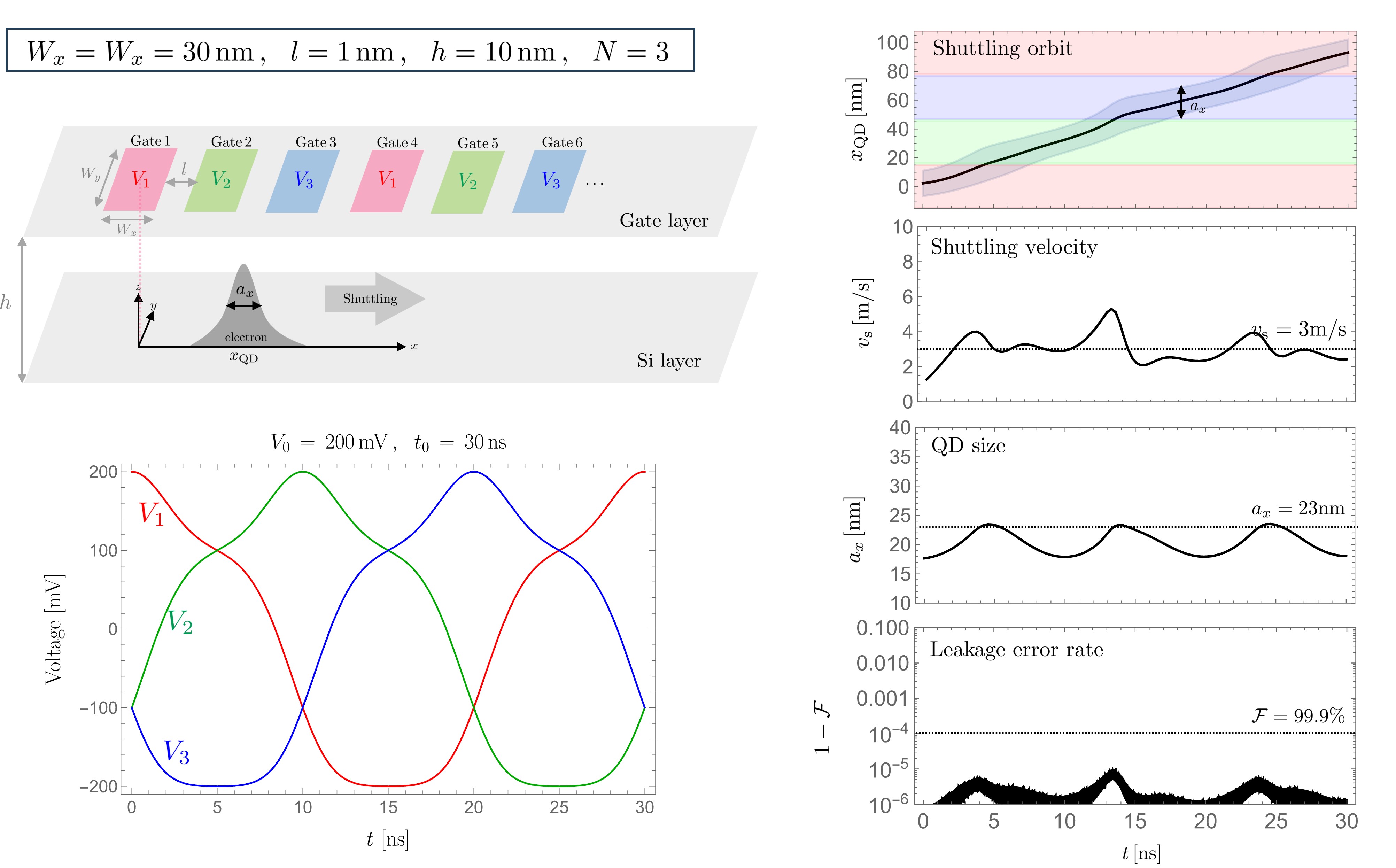}
  \caption{
   Simulation results of the conventional analog control shuttling under the same conditions as in Fig.~\ref{fig-sim-proposed}, except the applied voltages are replaced by Eqs.~(\ref{eq:Vpulse}) with $V_0=200\,\mathrm{mV}$, $t_0=30\,\mathrm{ns}$, and $\tau/t_0=0.1$. The plotted quantities match those in Fig.~\ref{fig-sim-proposed}.}
  \label{fig-sim-conventional}
\end{figure*}
We show the results of the simulation of conventional method in Fig. \ref{fig-sim-conventional}.
The setup is the same with figure \ref{fig-sim-proposed}. We observe that the leakage rate $1-\mathcal{F}$ remains below $10^{-4}$, which is comparable with our proposed method (shown in figure \ref{fig-sim-proposed}).

\section{Circuit-simulation-based robustness evaluation}
\label{app:circuitsim}

\begin{figure*}[t]
  \centering
  \includegraphics[width=18cm]{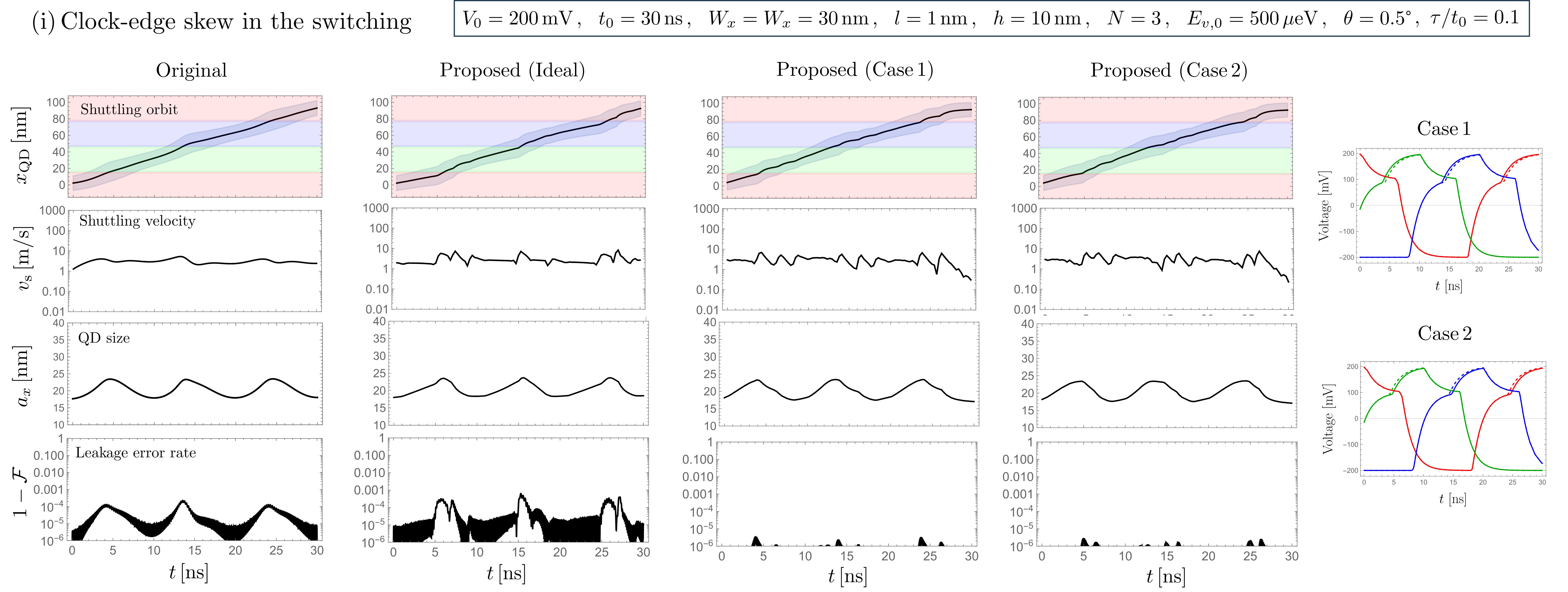}\\[0.5cm]
  \includegraphics[width=18cm]{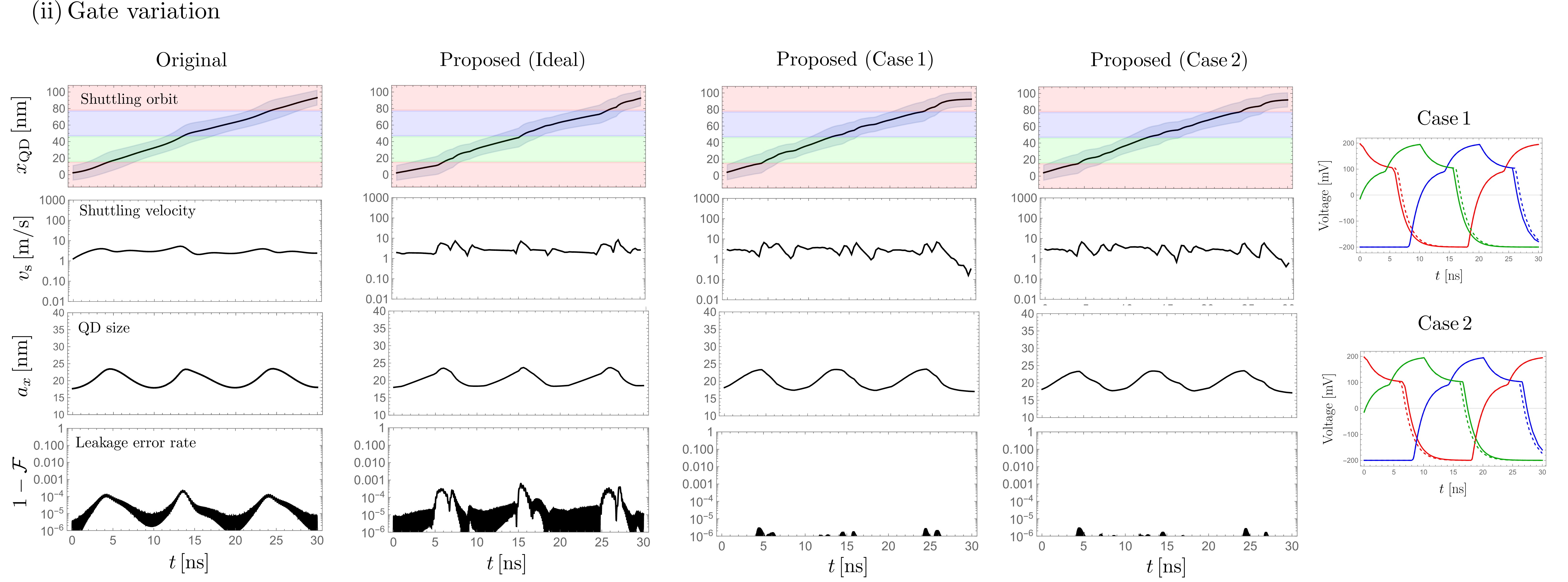}\\[0.5cm]
  \includegraphics[width=18cm]{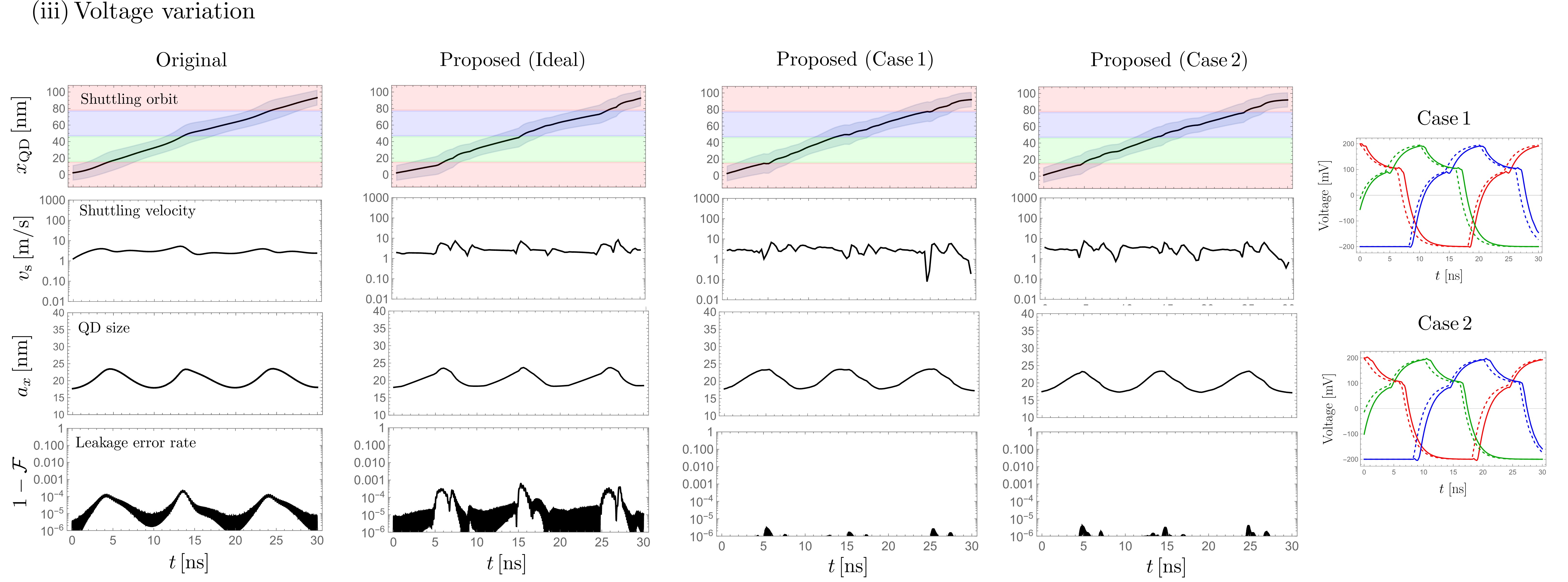}
  \caption{Examples of circuit simulations of the proposed shuttling method. The left two colums of the top, middle, and bottom figures show the result of the case with (i) clock-edge skew in the switching, (ii) gate variations, and (iii) voltage variation.}
  \label{fig-sim}
\end{figure*}

Based on the structure shown in Fig.~\ref{fig-structure}, we performed circuit simulations to analyze the impact of waveform variations that may occur in an actual circuit. 
We use Cadence Spectre simulator for the circuit simulation.
The results of these simulations are presented in Figs.~\ref{fig-sim}. In this analysis, the following parameter values were used:
$W_x = W_y = 30\,\mathrm{nm},\quad l = 1\,\mathrm{nm},\quad N = 3,\quad V_0 = 200\,\mathrm{mV},\quad 1/f = 30\,\mathrm{ns},\quad \tau f = 0.1,\quad$$E_{v,0} = 500\,\mu\mathrm{eV},$ and $\theta = 0.3^\circ$.

As representative sources of waveform variation, we considered the following factors: (i) Timing generation edge skew (top panels of Fig.~\ref{fig-sim}), (ii) Gate-to-gate variability in characteristics (center panels of Fig.~\ref{fig-sim}), and  (iii) Variability among voltage selection channels (bottom panels of Fig.~\ref{fig-sim}).
For each factor, two variation patterns, as indicated by the solid lines in the rightmost column, have been analyzed (the dotted lines indicate the voltages in the absence of any variation). In each figure, the first two columns from the right display the effects of these variations, while the leftmost column shows the results for the conventional approach and the second column from the left shows the results for the proposed method in the absence of variations. The time dependence of the shuttling orbit $(x_{\rm{QD}})$, velocity $(v_{\rm{s}})$, the quantum dot size $(a_x)$ and the leakage error rate $(1-\mathcal{F})$ are respectively shown from top to bottom figures. 
The simulation results indicate that, in all cases, no degradation in fidelity is observed, confirming robust performance against these
waveform variations.

\bibliography{ref}

\end{document}